\documentclass{PoS}
\usepackage{lineno}  % for line numbering during review
\usepackage{mciteplus}
\usepackage{graphicx}
\usepackage{ifthen} % for conditional statements
\usepackage{amsmath} % Adds a large collection of math symbols
\usepackage{amssymb}
\usepackage{amsfonts}
\usepackage{upgreek}
\usepackage{cite} % Allows for ranges in citations
\usepackage{mciteplus}
\usepackage{xspace}
\usepackage{pstricks}
\usepackage{tablefootnote}
\newboolean{uprightparticles}
\setboolean{uprightparticles}{false} %True for upright particle symbols
\newboolean{inbibliography}
\setboolean{inbibliography}{false} %True once you enter the bibliography
\newboolean{articletitles}
\setboolean{articletitles}{false} % False removes titles in references
\def\lhcb {\mbox{LHCb}\xspace}

\def\dzero  {\mbox{D0}\xspace}

\def\lhc    {\mbox{LHC}\xspace}

\def\MagUp {\mbox{\em Mag\kern -0.05em Up}\xspace}

\ifthenelse{\boolean{uprightparticles}}%
{

 \def\Pmu         {\ensuremath{\upmu}\xspace}

 \def\Ppi         {\ensuremath{\uppi}\xspace}                 
                  
 \def\Prho        {\ensuremath{\uprho}\xspace}

 \def\Pphi        {\ensuremath{\upphi}\xspace}

 \def\Ppsi        {\ensuremath{\uppsi}\xspace}

 \def\PDelta      {\ensuremath{\Delta}\xspace}                 
 \def\PXi      {\ensuremath{\Xi}\xspace}                 
 \def\PLambda      {\ensuremath{\Lambda}\xspace}                 
 \def\PSigma      {\ensuremath{\Sigma}\xspace}                 
 \def\POmega      {\ensuremath{\Omega}\xspace}                 
 \def\PUpsilon      {\ensuremath{\Upsilon}\xspace}                 
 
 \def\PB      {\ensuremath{\mathrm{B}}\xspace}                 
                  
 \def\PD      {\ensuremath{\mathrm{D}}\xspace}

 \def\PJ      {\ensuremath{\mathrm{J}}\xspace}                 
 \def\PK      {\ensuremath{\mathrm{K}}\xspace}

 \def\PW      {\ensuremath{\mathrm{W}}\xspace}

 \def\Pb      {\ensuremath{\mathrm{b}}\xspace}                 
 \def\Pc      {\ensuremath{\mathrm{c}}\xspace}                 
 \def\Pd      {\ensuremath{\mathrm{d}}\xspace}

 \def\Pi      {\ensuremath{\mathrm{i}}\xspace}

 \def\Pp      {\ensuremath{\mathrm{p}}\xspace}

 \def\Ps      {\ensuremath{\mathrm{s}}\xspace}                 
 \def\Pt      {\ensuremath{\mathrm{t}}\xspace}                 
 \def\Pu      {\ensuremath{\mathrm{u}}\xspace}

}
{

 \def\Pmu         {\ensuremath{\mu}\xspace}

 \def\Ppi         {\ensuremath{\pi}\xspace}                 
                  
 \def\Prho        {\ensuremath{\rho}\xspace}

 \def\Pphi        {\ensuremath{\phi}\xspace}

 \def\Ppsi        {\ensuremath{\psi}\xspace}                 
                  
 \mathchardef\PDelta="7101
 \mathchardef\PXi="7104
 \mathchardef\PLambda="7103
 \mathchardef\PSigma="7106
 \mathchardef\POmega="710A
 \mathchardef\PUpsilon="7107
                  
 \def\PB      {\ensuremath{B}\xspace}                 
                  
 \def\PD      {\ensuremath{D}\xspace}

 \def\PJ      {\ensuremath{J}\xspace}                 
 \def\PK      {\ensuremath{K}\xspace}

 \def\PW      {\ensuremath{W}\xspace}

 \def\Pb      {\ensuremath{b}\xspace}                 
 \def\Pc      {\ensuremath{c}\xspace}                 
 \def\Pd      {\ensuremath{d}\xspace}

 \def\Pi      {\ensuremath{i}\xspace}

 \def\Pp      {\ensuremath{p}\xspace}

 \def\Ps      {\ensuremath{s}\xspace}                 
 \def\Pt      {\ensuremath{t}\xspace}                 
 \def\Pu      {\ensuremath{u}\xspace}

}

\makeatletter
\ifcase \@ptsize \relax% 10pt
  \newcommand{\miniscule}{\@setfontsize\miniscule{4}{5}}% \tiny: 5/6
\or% 11pt
  \newcommand{\miniscule}{\@setfontsize\miniscule{5}{6}}% \tiny: 6/7
\or% 12pt
  \newcommand{\miniscule}{\@setfontsize\miniscule{5}{6}}% \tiny: 6/7
\fi
\makeatother

\DeclareRobustCommand{\optbar}[1]{\shortstack{{\miniscule (\rule[.5ex]{1.25em}{.18mm})}
  \\ [-.7ex] $#1$}}

\def\mup        {{\ensuremath{\Pmu^+}}\xspace}
\def\mun        {{\ensuremath{\Pmu^-}}\xspace} % muon negative (\mum is taken)
\def\mumu       {{\ensuremath{\Pmu^+\Pmu^-}}\xspace}

\def\W      {{\ensuremath{\PW}}\xspace}

\def\uquark    {{\ensuremath{\Pu}}\xspace}

\def\dquark    {{\ensuremath{\Pd}}\xspace}

\def\squark    {{\ensuremath{\Ps}}\xspace}
\def\squarkbar {{\ensuremath{\overline \squark}}\xspace}

\def\cquark    {{\ensuremath{\Pc}}\xspace}
\def\cquarkbar {{\ensuremath{\overline \cquark}}\xspace}

\def\bquark    {{\ensuremath{\Pb}}\xspace}
\def\bquarkbar {{\ensuremath{\overline \bquark}}\xspace}

\def\tquark    {{\ensuremath{\Pt}}\xspace}

\def\pion   {{\ensuremath{\Ppi}}\xspace}
\def\piz    {{\ensuremath{\pion^0}}\xspace}

\def\pip    {{\ensuremath{\pion^+}}\xspace}
\def\pim    {{\ensuremath{\pion^-}}\xspace}
\def\pipm   {{\ensuremath{\pion^\pm}}\xspace}

\def\rhomeson {{\ensuremath{\Prho}}\xspace}
\def\rhoz     {{\ensuremath{\rhomeson^0}}\xspace}
\def\rhop     {{\ensuremath{\rhomeson^+}}\xspace}
\def\rhom     {{\ensuremath{\rhomeson^-}}\xspace}

\def\kaon    {{\ensuremath{\PK}}\xspace}
  \def\Kbar    {{\kern 0.2em\overline{\kern -0.2em \PK}{}}\xspace}

\def\KorKbar    {\kern 0.18em\optbar{\kern -0.18em K}{}\xspace}
\def\Kz      {{\ensuremath{\kaon^0}}\xspace}

\def\Kp      {{\ensuremath{\kaon^+}}\xspace}
\def\Km      {{\ensuremath{\kaon^-}}\xspace}
\def\Kpm     {{\ensuremath{\kaon^\pm}}\xspace}

\def\KS      {{\ensuremath{\kaon^0_{\rm\scriptscriptstyle S}}}\xspace}

\def\Kstarzb {{\ensuremath{\Kbar{}^{*0}}}\xspace}
\def\Kstar   {{\ensuremath{\kaon^*}}\xspace}

\def\Dbar    {{\kern 0.2em\overline{\kern -0.2em \PD}{}}\xspace}
\def\D       {{\ensuremath{\PD}}\xspace}

\def\DorDbar    {\kern 0.18em\optbar{\kern -0.18em D}{}\xspace}
\def\Dz      {{\ensuremath{\D^0}}\xspace}
\def\Dzb     {{\ensuremath{\Dbar{}^0}}\xspace}
\def\Dp      {{\ensuremath{\D^+}}\xspace}

\def\Dstar   {{\ensuremath{\D^*}}\xspace}

\def\Ds      {{\ensuremath{\D^+_\squark}}\xspace}
\def\Dsp     {{\ensuremath{\D^+_\squark}}\xspace}
\def\Dsm     {{\ensuremath{\D^-_\squark}}\xspace}
\def\Dspm    {{\ensuremath{\D^{\pm}_\squark}}\xspace}

\def\B       {{\ensuremath{\PB}}\xspace}
\def\Bbar    {{\ensuremath{\kern 0.18em\overline{\kern -0.18em \PB}{}}}\xspace}

\def\BorBbar    {\kern 0.18em\optbar{\kern -0.18em B}{}\xspace}
\def\Bz      {{\ensuremath{\B^0}}\xspace}
\def\Bzb     {{\ensuremath{\Bbar{}^0}}\xspace}
\def\Bu      {{\ensuremath{\B^+}}\xspace}
\def\Bub     {{\ensuremath{\B^-}}\xspace}
\def\Bp      {{\ensuremath{\Bu}}\xspace}
\def\Bm      {{\ensuremath{\Bub}}\xspace}
\def\Bpm     {{\ensuremath{\B^\pm}}\xspace}

\def\Bd      {{\ensuremath{\B^0}}\xspace}
\def\Bs      {{\ensuremath{\B^0_\squark}}\xspace}

\def\Bdb     {{\ensuremath{\Bbar{}^0}}\xspace}

\def\jpsi     {{\ensuremath{{\PJ\mskip -3mu/\mskip -2mu\Ppsi\mskip 2mu}}}\xspace}

  \def\Y#1S{\ensuremath{\PUpsilon{(#1S)}}\xspace}% no space before {...}!

\def\proton      {{\ensuremath{\Pp}}\xspace}

\def\Lz          {{\ensuremath{\PLambda}}\xspace}
\def\Lbar        {{\ensuremath{\kern 0.1em\overline{\kern -0.1em\PLambda}}}\xspace}
\def\LorLbar    {\kern 0.18em\optbar{\kern -0.18em \PLambda}{}\xspace}

\def\Lb      {{\ensuremath{\Lz^0_\bquark}}\xspace}

\def\Lc      {{\ensuremath{\Lz^+_\cquark}}\xspace}

\newcommand{\decay}[2]{\ensuremath{#1\!\to #2}\xspace}         % {\Pa}{\Pb \Pc}

\def\to                 {\ensuremath{\rightarrow}\xspace}

\def\CP                {{\ensuremath{C\!P}}\xspace}

\def\Vud  {{\ensuremath{V_{\uquark\dquark}}}\xspace}

\def\Vtd  {{\ensuremath{V_{\tquark\dquark}}}\xspace}

\def\Vcs  {{\ensuremath{V_{\cquark\squark}}}\xspace}
\def\Vts  {{\ensuremath{V_{\tquark\squark}}}\xspace}
\def\Vub  {{\ensuremath{V_{\uquark\bquark}}}\xspace}
\def\Vcb  {{\ensuremath{V_{\cquark\bquark}}}\xspace}

\def\Vubs  {{\ensuremath{V_{\uquark\bquark}^\ast}}\xspace}
\def\Vcbs  {{\ensuremath{V_{\cquark\bquark}^\ast}}\xspace}
\def\Vtbs  {{\ensuremath{V_{\tquark\bquark}^\ast}}\xspace}

\def\AT#1     {\ensuremath{A_{\mathrm{T}}^{#1}}\xspace}           % 2

\def\C#1      {\ensuremath{\mathcal{C}_{#1}}\xspace}                       % 9
\def\Cp#1     {\ensuremath{\mathcal{C}_{#1}^{'}}\xspace}                    % 7
\def\Ceff#1   {\ensuremath{\mathcal{C}_{#1}^{\mathrm{(eff)}}}\xspace}        % 9  
\def\Cpeff#1  {\ensuremath{\mathcal{C}_{#1}^{'\mathrm{(eff)}}}\xspace}       % 7
\def\Ope#1    {\ensuremath{\mathcal{O}_{#1}}\xspace}                       % 2
\def\Opep#1   {\ensuremath{\mathcal{O}_{#1}^{'}}\xspace}                    % 7

\newcommand{\tev}{\ifthenelse{\boolean{inbibliography}}{\ensuremath{~T\kern -0.05em eV}\xspace}{\ensuremath{\mathrm{\,Te\kern -0.1em V}}}\xspace}
\newcommand{\gev}{\ensuremath{\mathrm{\,Ge\kern -0.1em V}}\xspace}
\newcommand{\mev}{\ensuremath{\mathrm{\,Me\kern -0.1em V}}\xspace}
\newcommand{\kev}{\ensuremath{\mathrm{\,ke\kern -0.1em V}}\xspace}
\newcommand{\ev}{\ensuremath{\mathrm{\,e\kern -0.1em V}}\xspace}
\newcommand{\gevc}{\ensuremath{{\mathrm{\,Ge\kern -0.1em V\!/}c}}\xspace}
\newcommand{\mevc}{\ensuremath{{\mathrm{\,Me\kern -0.1em V\!/}c}}\xspace}
\newcommand{\gevcc}{\ensuremath{{\mathrm{\,Ge\kern -0.1em V\!/}c^2}}\xspace}
\newcommand{\gevgevcccc}{\ensuremath{{\mathrm{\,Ge\kern -0.1em V^2\!/}c^4}}\xspace}
\newcommand{\mevcc}{\ensuremath{{\mathrm{\,Me\kern -0.1em V\!/}c^2}}\xspace}

\def\mub{\ensuremath{{\rm \,\upmu b}}\xspace}

\def\invpb {\ensuremath{\mbox{\,pb}^{-1}}\xspace}

\def\invfb   {\ensuremath{\mbox{\,fb}^{-1}}\xspace}

\newcommand{\chisqip}{\ensuremath{\chi^2_{\rm IP}}\xspace}

\def\gsim{{~\raise.15em\hbox{$>$}\kern-.85em
          \lower.35em\hbox{$\sim$}~}\xspace}
\def\lsim{{~\raise.15em\hbox{$<$}\kern-.85em
          \lower.35em\hbox{$\sim$}~}\xspace}

\def\pt         {\mbox{$p_{\rm T}$}\xspace}

\def\degrees{\ensuremath{^{\circ}}\xspace}

\def\rad{\ensuremath{\rm \,rad}\xspace}

\def\tell1  {TELL1\xspace}
\def\ukl1   {UKL1\xspace}

\def\ellell{\ensuremath{\ell^+\ell^-}\xspace}
\usepackage{dcolumn}
\newcolumntype{C}{>{$}c<{$}}
\newcolumntype{R}{>{$}r<{$}}
\newcolumntype{L}{>{$}l<{$}}
\newcommand{\aerr}[2]{{\:}^{+{\:}#1}_{-{\:}#2}}%
\usepackage{wrapfig}
\usepackage{cite}
\newcommand{\skipit}[1]{} % do nothing!
\title{\CP violation and CKM studies (and first LHCb Run II results)}

\ShortTitle{\CP violation and CKM studies}

\author{\speaker{Patrick Koppenburg}\\
        Nikhef and CERN\\
        E-mail: \email{patrick.koppenburg@cern.ch}}

\def\theabstract{The LHC is the new \bquark-hadron factory and will be dominating flavour physics until
the start of Belle II, and beyond in many decay modes. While the \B factories and Tevatron experiments 
are still analysing their data, ATLAS, CMS and LHCb
are producing interesting new results in \CP violation and rare decays, that set strong constraints 
on models beyond that SM and exhibit some discrepancies with the SM
predictions. The LHCb collaboration used the LHC 50\:ns ramp-up period of July 2015 to measure the double-differential \jpsi, \jpsi-from-\bquark-hadron and charm cross-sections at $\sqrt{s}=13\:\tev$. Both measurements were performed directly on triggered candidates using a reduced data format that does not require offline processing.}
\abstract{\theabstract}

\FullConference{The European Physical Society Conference on High Energy Physics\\
		22--29 July 2015\\
		Vienna, Austria}

\begin{document}
%%%%%%%%%%%%%%%%%%%%%%%%%%%%%%%%%%%%%%%%%%%%%%%%%%%%%%%%%
%%%%%%%%%%%%%%%%%%%%%%%%%%%%%%%%%%%%%%%%%%%%%%%%%%%%%%%%%
\section{Introduction}
The first run of the Large Hadron Collider with 7 and 8\:\tev $pp$ collisions has allowed
to discover the Higgs boson~\cite{Aad:2012tfa,*Chatrchyan:2012ufa}, but not to find any hint of 
the existence of other
new particles. Neither supersymmetry nor any other sign of new
physics has popped out of the LHC. This situation may change during Run II
with a higher centre-of-mass energy of 13\:\tev. At the same time
it is worth investigating what the data at the lower energies involved in
the decays of \bquark hadrons can tell us about new physics.

Studies of \CP violation in heavy flavour decays are both sensitive 
to the above-mentioned
high mass scales and to potential new phases beyond the phase of the CKM matrix. 
Also of particular interest are rare decays that are strongly suppressed in the Standard Model (SM), where new physics amplitudes could be sizeable~\cite{Karim}.

\section{The Kobayashi-Maskawa mechanism}\label{sec:ckmhistory}
Flavour physics has played a prominent  role in the development of the SM. As an example, one of the most notable predictions made in this context was that of the existence of a third quark generation, in a famous paper of 1973 by Makoto Kobayashi and Toshihide Maskawa~\cite{Kobayashi:1973fv}. This work won them the Nobel Price in Physics in 2008 ``for the discovery of the origin of the broken symmetry which predicts the existence of at least three families of quarks in nature''. Kobayashi and Maskawa extended the Cabibbo~\cite{Cabibbo:1963yz} (with only $u$, $d$, and $s$ quarks) and the Glashow-Iliopoulos-Maiani~\cite{Glashow:1970gm} (GIM, including also the $c$ quark) mechanisms, pointing out that $C\!P$ violation could be incorporated into the emerging picture of the SM if six quarks were present. This is commonly referred to as the Kobayashi-Maskawa (KM) mechanism. It must be emphasised that at that time only hadrons made of the three lighter quarks had been observed. An experimental revolution took place in 1974, when a new state containing the $c$ quark was discovered almost simultaneously at Brookhaven~\cite{Aubert:1974js} and SLAC~\cite{Augustin:1974xw}. Then, the experimental observations of the $b$~\cite{Herb:1977ek} and $t$~\cite{Abe:1995hr} quarks were made at FNAL in 1977 and 1995, respectively. 

The idea of Kobayashi and Maskawa, formalised in the so-called Cabibbo-Kobayashi-Maskawa (CKM) quark mixing matrix, was then included in the SM by the beginning of the 1980's. The phenomenon of $C\!P$ violation, first revealed in 1964 using decays of neutral kaons~\cite{Christenson:1964fg}, was elegantly accounted for as an irreducible complex phase within the CKM matrix. The experimental proof of the validity of the KM mechanism and the precise measurement of the value of the \CP-violating phase soon became questions of paramount importance.

\subsection{The CKM matrix}\label{sec:CKM}\label{sec:CKMdefinition}
In the SM, charged-current interactions of quarks are described by the Lagrangian
\begin{equation*}\label{eq:chargeInteraction}
  \mathcal{L}_{W^{\pm}} = - \frac{g}{\sqrt{2}}\overline{U}_{i}\gamma^{\mu}\frac{1-\gamma^5}{2}\left(V_{\rm CKM} \right )_{ij} D_{j} W_{\mu}^{+} + h.c.,
\end{equation*}
where $g$ is the electroweak coupling constant and $V_{\rm CKM}$ is the CKM matrix
\begin{equation*}\label{eq:ckmMatrix}
  V_{\rm CKM} = \left( 
    \begin{array}{lcr}
      V_{ud} & V_{us} & V_{ub} \\
      V_{cd} & V_{cs} & V_{cb} \\
      V_{td} & V_{ts} & V_{tb}
    \end{array}
  \right),
\end{equation*}
originating from the misalignment in flavour space of the up and down components of the $SU(2)_L$ quark doublet of the SM. The $V_{ij}$ matrix elements represent the couplings between up-type quarks $U_i=(u,\,c,\,t)$ and down-type quarks $D_j=(d,\,s,\,b)$.

\subsection{Wolfenstein parameterisation}\label{sec:CKMWolfenstein}

Experimental information leads to the evidence that transitions within the same generation are characterised by $V_{\rm CKM}$ elements of $\mathcal{O}(1)$. Instead, those between the first and second generations are suppressed by a factor $\mathcal{O}(10^{-1})$; those between the second and third generations by a factor $\mathcal{O}(10^{-2})$; and those between the first and third generations by a factor $\mathcal{O}(10^{-3})$. 
The CKM matrix can be re-written as a power expansion of the parameter $\lambda$ (which corresponds to $\sin{\theta_{C}}$)~\cite{Wolfenstein:1983yz}
\begin{equation*}\label{eq:ckmWolfenstein4}
\resizebox{1\hsize}{!}{$
V_{\rm CKM} = \left(
  \begin{array}{ccc}
    1-\frac{1}{2}\lambda^{2}-\frac{1}{8}\lambda^{4}                                   & \lambda                  & A\lambda^{3}\left(\rho -i\eta\right) \\
    -\lambda +\frac{1}{2}A^{2}\lambda^{5}\left[1-2(\rho +i\eta)\right]        & 1-\frac{1}{2}\lambda^{2}-\frac{1}{8}\lambda^{4}(1+4A^{2})     & A\lambda^{2}                                     \\
    A\lambda^{3}\left[1-(\rho +i\eta)\left(1-\frac{1}{2}\lambda^{2}\right)\right]~ & ~-A\lambda^{2}+\frac{1}{2}A\lambda^{4}\left[1-2(\rho +i\eta)\right]~        & ~1-\frac{1}{2}A^{2}\lambda^{4}
  \end{array}
\right),
$}
\end{equation*}
which is valid up to $\mathcal{O}\left(\lambda^{6}\right)$. With this parameterisation, the CKM matrix is complex, and hence \CP violation is allowed for, if and only if $\eta$ differs from zero.
To lowest order the Jarlskog parameter~\cite{Jarlskog:1985ht} is
\begin{equation*}\label{eq:jarlskogValueWolf}
J_{C\!P} \equiv \left|\Im\left(V_{i\alpha} V_{j\beta} V_{i\beta}^{*}V_{j\alpha}^{*}\right)\right| = \lambda^{6}A^{2}\eta, \qquad \left( i\neq j, \alpha\neq\beta\right),
\end{equation*}
and, as expected, is directly related to the \CP-violating parameter $\eta$.
\subsection{The unitarity triangle}\label{sec:UT}

\begin{figure}[tb]
  \begin{center}
    \includegraphics[width=1\textwidth]{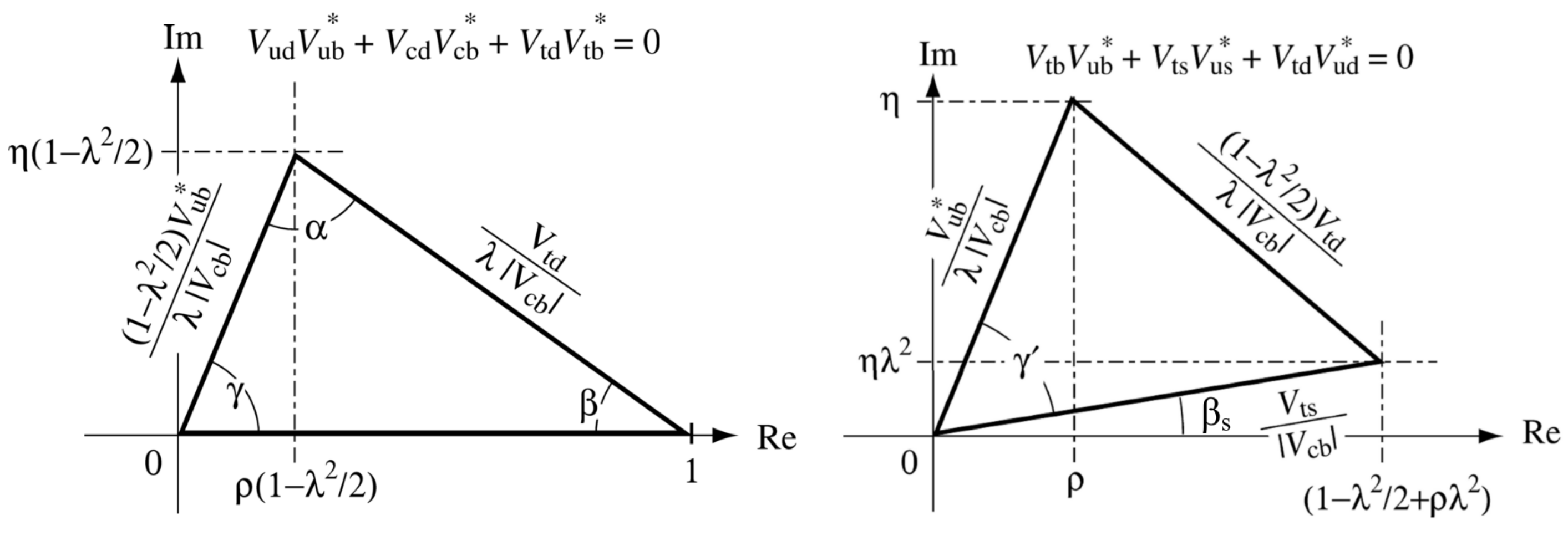}
  \end{center}
  \caption{Representation in the complex plane of the non-squashed triangles obtained from the off-diagonal unitarity relations of the CKM matrix.}\label{fig:unitaryTriangle}
\end{figure}

The unitarity condition of the CKM matrix, $V_{\rm CKM}V_{\rm CKM}^{\dagger} = V_{\rm CKM}^{\dagger}V_{\rm CKM} = \mathbb{I}$, leads to a set of 12 equations: 6 for diagonal terms and 6 for off-diagonal terms. In particular, the equations for the off-diagonal terms can be represented as triangles in the complex plane, all characterised by the same area $J_{C\!P}/2$. Only two out of these six triangles have sides of the same order of magnitude, $\mathcal{O}(\lambda^{3})$:
\begin{eqnarray*}
%  \underbrace{V_{ud}V_{us}^{*}}_{\mathcal{O}(\lambda)}+\underbrace{V_{cd}V_{cs}^{*}}_{\mathcal{O}(\lambda)}+\underbrace{V_{td}V_{ts}^{*}}_{\mathcal{O}(\lambda^{5})} & = & 0, \\
%  \underbrace{V_{us}V_{ub}^{*}}_{\mathcal{O}(\lambda^{4})}+\underbrace{V_{cs}V_{cb}^{*}}_{\mathcal{O}(\lambda^{2})}+\underbrace{V_{ts}V_{tb}^{*}}_{\mathcal{O}(\lambda^{2})} & = & 0, \\
  \underbrace{V_{ud}V_{ub}^{*}}_{\mathcal{O}(\lambda^{3})}+\underbrace{V_{cd}V_{cb}^{*}}_{\mathcal{O}(\lambda^{3})}+\underbrace{V_{td}V_{tb}^{*}}_{\mathcal{O}(\lambda^{3})} & = & 0, \\
%  \underbrace{V_{ud}V_{cd}^{*}}_{\mathcal{O}(\lambda)}+\underbrace{V_{us}V_{cs}^{*}}_{\mathcal{O}(\lambda)}+\underbrace{V_{ub}V_{cb}^{*}}_{\mathcal{O}(\lambda^{5})} & = & 0, \\
%  \underbrace{V_{cd}V_{td}^{*}}_{\mathcal{O}(\lambda^{4})}+\underbrace{V_{cs}V_{ts}^{*}}_{\mathcal{O}(\lambda^{2})}+\underbrace{V_{cb}V_{tb}^{*}}_{\mathcal{O}(\lambda^{2})} & = & 0, \\
  \underbrace{V_{ud}V_{td}^{*}}_{\mathcal{O}(\lambda^{3})}+\underbrace{V_{us}V_{ts}^{*}}_{\mathcal{O}(\lambda^{3})}+\underbrace{V_{ub}V_{tb}^{*}}_{\mathcal{O}(\lambda^{3})} & = & 0.
\end{eqnarray*}
These two triangles in the complex plane are represented in Fig.~\ref{fig:unitaryTriangle}. In particular, the triangle defined by the former equation is commonly referred to as \emph{the} unitarity triangle (UT).
The sides of the UT are given by
\begin{equation*}\label{eq:utRB}
  R_{u} \equiv \left|\frac{V_{ud}V_{ub}^{*}}{V_{cd}V_{cb}^{*}}\right|  =  \sqrt{\bar{\rho}^{2}+\bar{\eta}^{2}},
\end{equation*}
\begin{equation*}
  R_{t} \equiv \left|\frac{V_{td}V_{tb}^{*}}{V_{cd}V_{cb}^{*}}\right| = \sqrt{\left(1-\bar{\rho}\right)^{2}+\bar{\eta}^{2}}, \label{eq:utRT}
\end{equation*}
where to simplify the notation the parameters $\bar{\rho}$ and $\bar{\eta}$, namely the coordinates in the complex plane of the only non-trivial apex of the UT, the others being (0, 0) and (1, 0), have been introduced. The angles of the UT are related to the CKM matrix elements as
\begin{equation*}
  \alpha\equiv \arg{\left(-\frac{V_{td}V_{tb}^{*}}{V_{ud}V_{ub}^{*}}\right)} = \arg{\left(-\frac{1-\bar{\rho}-i\bar{\eta}}{\bar{\rho}+i\bar{\eta}}\right)},\label{eq:alpha}
\end{equation*}
\begin{equation*}
  \beta \equiv\arg{\left(-\frac{V_{cd}V_{cb}^{*}}{V_{td}V_{tb}^{*}}\right)} = \arg{\left(\frac{1}{1-\bar{\rho}-i\bar{\eta}}\right)}, \label{eq:beta}
\end{equation*}
\begin{equation*}
  \gamma\equiv\arg{\left(-\frac{V_{ud}V_{ub}^{*}}{V_{cd}V_{cb}^{*}}\right)} = \arg{\left(\bar{\rho}+i\bar{\eta}\right)}. \label{eq:gamma}
\end{equation*}
The second non-squashed triangle has similar characteristics with respect to the UT. The apex is placed in the point $(\rho ,~\eta)$ and is tilted by an angle
\begin{equation*}\label{eq:betas}
  \beta_{s} \equiv \arg{\left(-\frac{V_{ts}V_{tb}^{*}}{V_{cs}V_{cb}^{*}}\right)} = \lambda^2 \eta + \mathcal{O}(\lambda^{4}).
\end{equation*}

%%%%%%%%%%%%%%%%%%%%%%%%%%%%%%%%%%%%%%%%%%%%%%%%%%%%%%%%%
\section{\boldmath\CP violation measurements}\label{Sec:CPV}
Owing to the legacy of the \B factories~\cite{Bevan:2014iga}, we have entered 
the era of precision tests of the CKM paradigm.%~\cite{Kobayashi:1973fv}.
Precise measurements of the angles of the unitarity triangle(s) are needed 
to search for new sources of \CP violation beyond the single phase of the CKM matrix.
%%%%%%%%%%%%%%%%%%%%%%%%%%%%%%%%%%%%%%%%%%%%%%%%%%%%%%%%%%%%
\begin{figure}[t]
  \def\LL{0.26\textwidth}
  \includegraphics[width=\textwidth]{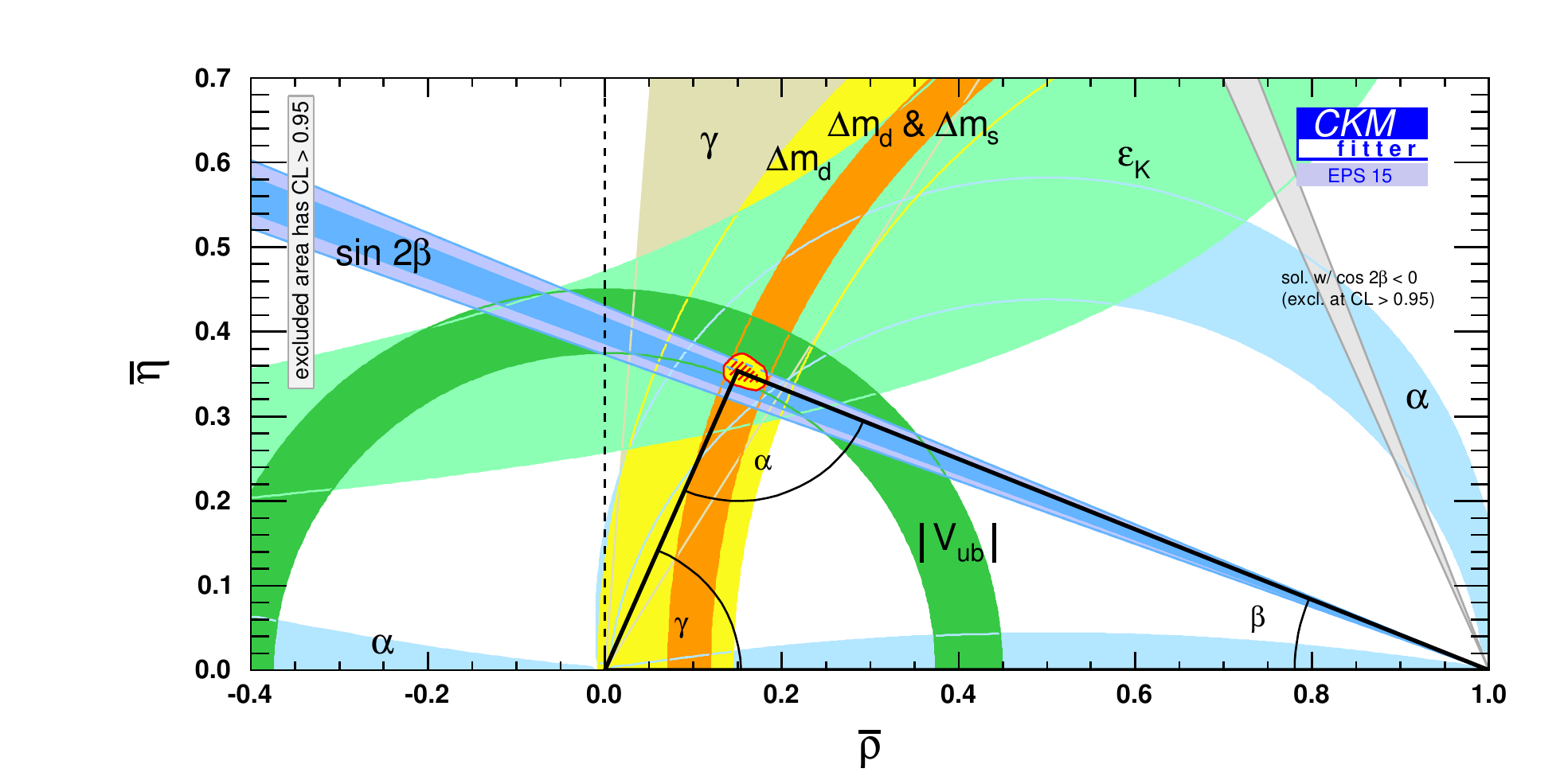}\hskip 0.04\textwidth%
  \caption{CKM unitarity triangle fit status in Summer 2015. Plot from Ref.~\cite{Charles:2004jd}.}\label{Fig:CKMFit}
\end{figure}
%%%%%%%%%%%%%%%%%%%%%%%%%%%%%%%%%%%%%%%%%%%%%%%%%%%%%%%%%%%%

In some cases, listed below, the same quantity---for instance the angle $\gamma$---can be measured in different ways and the results can be compared. In most cases though, the measurements serve as inputs to global fits of the unitarity triangle and the overall consistency is checked. The latest fit from the CKMFitter group is shown in Fig.~\ref{Fig:CKMFit}~\cite{Charles:2004jd}. Similar fits with slightly different assumptions are done by the UTFit collaboration~\cite{Bona:2006ah}. The overall consistency is good with the present uncertainties, though some tensions can be seen between the intersection of the \Vub circle and $\Delta m_s$ circles, and the $\sin2\beta$ diagonal. See below for more details.

%%%%%%%%%%%%%%%%%%%%%%%%%%%%%%%%%%%%%%%%%%%%%%%%%%%%%%%%%%%%%%%%%%%%%%%%%%%%%%%%%%%%%
\subsection{\boldmath\Bd mixing}\label{Sec:BdMix}
%%%%%%%%%%%%%%%%%%%%%%%%%%%%%%%%%%%%%%%%%%%%%%%%%%%%%%%%%%%%
\begin{figure}[b]
  \def\LL{0.23\textwidth}
  \includegraphics[height=\LL]{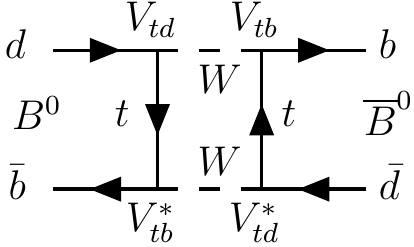}\hskip 0.1\textwidth%
  \includegraphics[height=\LL]{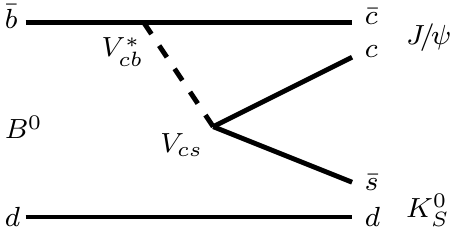}%\hskip 0.04\textwidth%
  \caption{Feynman diagrams of (left) the dominating SM contribution to \Bd oscillations,
    and (right) the decay \decay{\Bd}{\jpsi\KS}.}\label{Fig:BdMix}\label{Fig:BdJpsiKs}
\end{figure}
%%%%%%%%%%%%%%%%%%%%%%%%%%%%%%%%%%%%%%%%%%%%%%%%%%%%%%%%%%%%
The mode which allowed for the first observation of \CP violation in \B decays is 
\decay{\Bd}{\jpsi\KS}. It allows the measurements of $\varphi_d$, the relative phase 
between the decay of the \Bz meson to {\jpsi\KS} and that of the oscillation of \Bd to \Bdb
followed by the decay \decay{\Bzb}{\jpsi\KS}. In the SM it is equal to $2\beta$~\cite{Bigi:1983cj}
and thus the amplitude of the oscillation is a measurement of $\sin2\beta$.
The diagrams corresponding to the dominating
SM amplitudes are shown in Fig.~\ref{Fig:BdMix}.
The \B factories were optimised for its measurement\cite{Bevan:2014iga,Aubert:2009aw,*Adachi:2012et}
and determined~\cite{HFAG}
 \begin{equation*}
\sin2\beta^\text{Measured} = 0.682 \pm 0.019,
\end{equation*}
which defines the most precise constraint on the unitarity triangle (Fig.~\ref{Fig:CKMFit}).
Recently LHCb joined the effort, publishing their first measurement of the time dependent \CP asymmetry 
in the decay \decay{\Bd}{\jpsi\KS} \cite{LHCb-PAPER-2015-004} with an uncertainty competitive with the individual measurements from the \B factories. With the upcoming LHC Run II, this will allow for further reducing the uncertainties.

The measured value is slightly lower than the expectation from all other constraints on the UT~\cite{Charles:2015gya}
\begin{equation*}
\sin2\beta^\text{Expected} = 0.771 \aerr{0.017}{0.041}.
\end{equation*}

This $2\sigma$ tension could be due to the so far neglected contribution from penguin topologies
in the decay  \decay{\Bd}{\jpsi\KS} or in other \decay{\bquark}{\cquark\cquarkbar\squark} decays to 
\CP eigenstates. This penguin pollution is described in more detail in Section~\ref{Sec:Penguins}.

Another approach is to use decays that are less (or differently) affected by such pollutions,
like decays to open charm. Babar and Belle have recently published their first joint paper,
about the decay \decay{\Bzb}{D_\CP^{(*)}h^0}~\cite{Abdesselam:2015gha}. The determined value of $\sin2\beta$ is 
consistent with expectations but the uncertainty does not yet allow to discriminate between the
observed and expected values mentioned above. This year LHCb also published a Dalitz plot analysis
of the decay \decay{\Bd}{\Dzb\pip\pim}, which will eventually allow a measurement of $\sin2\beta$~\cite{LHCb-PAPER-2014-070}.

The frequency of \Bd mixing $\Delta m_d$ also sets strong constraints on the Unitarity Triangle (UT)
side opposite to the angle $\gamma$, especially when combined with its \Bs counterpart $\Delta m_s$.
LHCb has recently presented a new preliminary measurement using \decay{\Bz}{D^{(*)-}\mup\nu_\mu X}.
A value of $(\Delta m_d)^\text{LHCb}=\left(503.6\pm2.0\pm1.3\right)\:\rm ns^{-1}$ is obtained
\cite{LHCb-CONF-2015-003,*LHCb-PAPER-2015-031}, which has a better precision than the HFAG
average~\cite{HFAG}, which including it becomes 
$(\Delta m_d)^\text{World}=\left(505.5 \pm 2.0\right)\:\rm ns^{-1}$. This result is not yet included in
the latest UT fits. 

%%%%%%%%%%%%%%%%%%%%%%%%%%%%%%%%%%%%%%%%%%%%%%%%%%%%%%%%%%%%%%%%%%%%%%%%%%%%%%%%%%%%%
\subsection{\boldmath\Bs mixing}\label{Sec:BsMix}
The LHC is often considered as a \Bs meson factory, owing to its large
cross-section and the unprecedented capabilities of the LHC experiments to precisely
resolve oscillations. 

%%%%%%%%%%%%%%%%%%%%%%%%%%%%%%%%%%%%%%%%%%%%%%%%%%%%%%%%%%%%
%\begin{wrapfigure}{r}{0.35\textwidth}
%\hskip 0.05\textwidth
%\begin{minipage}{0.3\textwidth}
%  \includegraphics[width=\textwidth]{Bs_mixing.pdf}
%  \caption{Feynman diagram of the dominating SM contribution to \Bs oscillations.}\label{Fig:BsMix}
%\end{minipage}
%\end{wrapfigure}
%%%%%%%%%%%%%%%%%%%%%%%%%%%%%%%%%%%%%%%%%%%%%%%%%%%%%%%%%%%%

\begin{figure}[tb]
\includegraphics[height=0.39\textwidth]{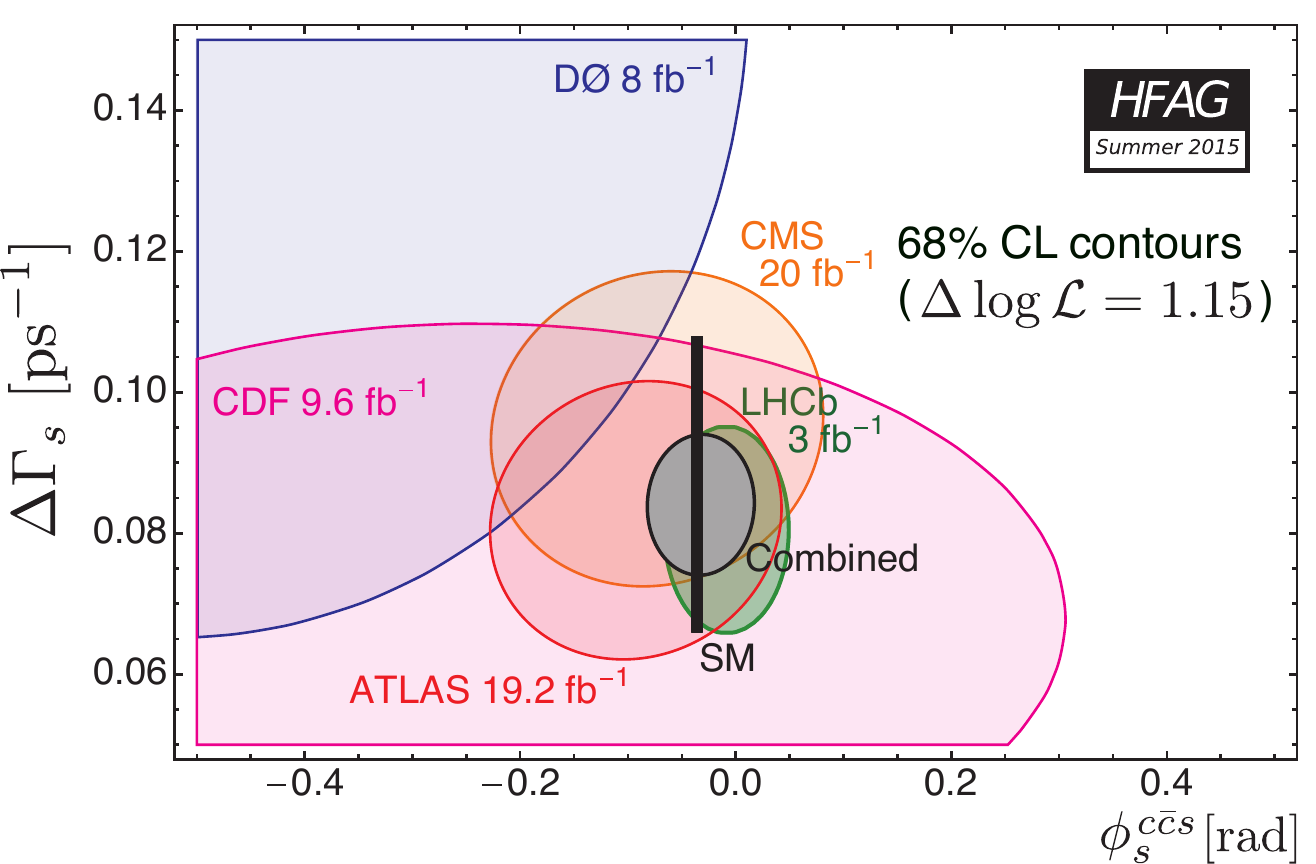}\hskip 0.02\textwidth 
\includegraphics[height=0.39\textwidth]{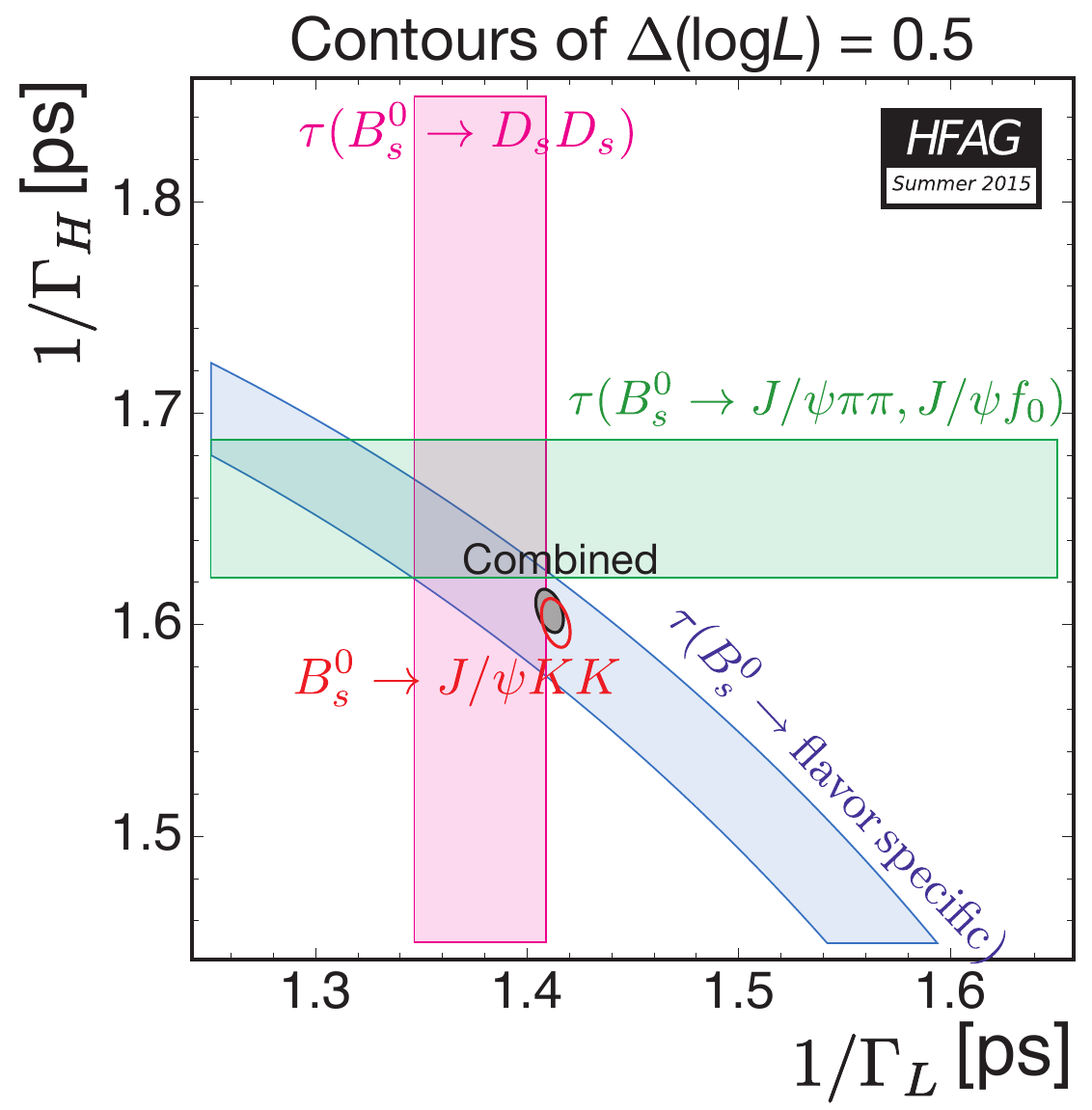}
\caption{ Constraints on (left) $\Delta\Gamma_s$ and 
  $\varphi_s^{\cquark\cquarkbar\squark}$ and  (right) $1/\Gamma_{L,H}$ from various decays and 
  experiments. Figure courtesy of HFAG~\cite{HFAG}. The SM predictions are from Refs.~\cite{Lenz:2011ti,*Lenz:2006hd,Charles:2004jd}.}
\label{Fig:2014-059}
\end{figure}

This opens the door to precision measurements of the \CP-violating
phase $\varphi_s^{\cquark\cquarkbar\squark}$, which is equal to 
$-2\beta_s\equiv-2\arg\left(-\Vts\Vtbs/\Vcs\Vcbs\right)=-0.0363\pm0.0013$ in the SM~\cite{Charles:2004jd}, neglecting sub-leading penguin contributions.
It was measured at the LHC using the flavour eigenstate \decay{\Bs}{\jpsi\phi} decay with \decay{\jpsi}{\mumu} and 
\decay{\phi}{\Kp\Km}~\cite{Aad:2014cqa,CMS-PAS-BPH-13-012,LHCb-PAPER-2013-002} and 
\decay{\Bs}{\jpsi\pip\pim}~\cite{LHCb-PAPER-2014-019}. Recently LHCb used the decay 
\decay{\Bs}{\jpsi\Kp\Km} for the first time in a polarisation-dependent way~\cite{LHCb-PAPER-2014-059}. Combined with \decay{\Bs}{\jpsi\pip\pim}, LHCb obtains 
$\varphi_s^{\cquark\cquarkbar\squark}=-0.010\pm 0.039$ rad. 
The decay mode \decay{\Bs}{\jpsi\phi} is also used by CMS~\cite{Khachatryan:2015nza}, 
ATLAS~\cite{Aad:2014cqa}, CDF~\cite{Aaltonen:2012ie} and \dzero~\cite{Abazov:2011ry}. The 
constraints on $\varphi_s^{\cquark\cquarkbar\squark}$ and the decay width difference 
$\Delta\Gamma_s=\Gamma_L-\Gamma_H$ are shown in Fig.~\ref{Fig:2014-059} (left).

The \Bs meson system has many features in common with that of the \Kz meson, with a heavy long-lived 
and a light short-lived eigenstate. The two lifetimes are shown in Fig~\ref{Fig:2014-059} (right).
The a priori unknown admixture of the two states contributing to a given non-flavour-specific decay
causes uncertainties in the measurement of branching fractions, for instance for 
the decay \decay{\Bs}{\mumu}, see Ref.~\cite{DeBruyn:2012wj,*DeBruyn:2012wk}. A precise determination
of the decay width difference is thus also important for the study of rare decays.

%%%%%%%%%%%%%%%%%%%%%%%
%%%%%%%%%%%%%%%%%%%%%%%
The same quantity was also measured with a fully hadronic final state using the decay
\decay{\Bs}{\Dsp\Dsm} with \decay{\Dspm}{\Kp\Km\pipm}, 
yielding $0.02\pm0.17\pm0.02$\:rad~\cite{LHCb-PAPER-2014-051}. 
The effective tagging power $\epsilon{ D}^2$ in excess of 
5\% for this channel is unprecedented at a hadron collider.

%%%%%%%%%%%%%%%%%%%%%%%%%%%%%%%%%%%%%%%%%%%%%%%%%%%%%%%%%%%%%%%%%%%%%%%%%%%%%%
\subsection{Flavour Tagging at the LHC}\label{Sec:Tagging}
This is just one example of analyses greatly indebted to improvements in the flavour tagging performance.
A better understanding of the underlying event and state-of-the-art multi-variate tagging methods allowed for 
increases in tagging performance up to a factor two between analyses on only 2011 data~\cite{LHCb-PAPER-2011-027}, and  those on the whole of Run II, see Table~\ref{Tab:Tagging}. These progresses are likely to continue. A good example
is a new tagging based on secondary charm hadrons recently introduced by LHCb~\cite{LHCb-PAPER-2015-027}.

%%%%%%%%%%%%%%%%%%%%%%%%%%%%%%%%%%%%%%%%%%%%%%%%%%%%%%%%%%%%%%%%%%%%%%%%%%%%%%%%%%%%%
\begin{table}[b]
\caption{Comparison of tagging performances $\epsilon D^2$ for selected time dependent \CP violation measurements at the LHC.}\label{Tab:Tagging}
\vskip 0.5em\centering
\begin{tabular}{ll|ccc}
  Experiment & Decay & 2011 & Run I & Improvement \\
  \hline
    LHCb & \decay{\Bs}{\phi\phi} & 3.29\%~\cite{LHCb-PAPER-2013-007} & 5.38\%~\cite{LHCb-PAPER-2014-026} & +64\%  \\
     & \decay{\Bs}{\Ds\Ds} &  & 5.33\%~\cite{LHCb-PAPER-2014-051}  \\
%    LHCb & \decay{\Bs}{\Ds\Km} & 5.07 &  &  & \LHCbShortCitation[\tiny\cyan]{LHCb-PAPER-2014-038} \\
     & \decay{\Bs}{\jpsi\Kp\Km} & 3.13\%~\cite{LHCb-PAPER-2013-002} & 3.73\%~\cite{LHCb-PAPER-2014-059} & +19\%  \\
     & \decay{\Bs}{\jpsi\pip\pim} & 2.43\%~\cite{LHCb-PAPER-2013-002} & 3.89\%~\cite{LHCb-PAPER-2014-019} & +60\%  \\
     & \decay{\Bd}{\jpsi\KS} & 2.38\%~\cite{LHCb-PAPER-2012-035} & 3.03\%~\cite{LHCb-PAPER-2015-004} & +27\%  \\
    \hline
    ATLAS & \decay{\Bs}{\jpsi\Pphi} & 1.45\%~\cite{Aad:2014cqa} & 1.49\%~\cite{Atlas:JpsiPhi} & \phantom{0}+3\% \\
    \hline
    CMS & \decay{\Bs}{\jpsi\Pphi} & & 1.31\%~\cite{Khachatryan:2015nza} & +35\%\tablefootnote{With respect to the preliminary result~\cite{CMS-PAS-BPH-13-012}.}  \\
\end{tabular}
\end{table}
%%%%%%%%%%%%%%%%%%%%%%%%%%%%%%%%%%%%%%%%%%%%%%%%%%%%%%%%%%%%%%%%%%%%%%%%%%%%%%%%%%%%%
%%%%%%%%%%%%%%%%%%%%%%%%%%%%%%%%%%%%%%%%%%%%%%%%%%%%%%%%%%%%%%%%%%%%%%%%%%%%%%%%%%%%%
\subsection{Penguin Topologies}\label{Sec:Penguins}
%%%%%%%%%%%%%%%%%%%%%%%%%%%%%%%%%%%%%%%%%%%%%%%%%%%%%%%%%%%%
\begin{wrapfigure}{r}{0.4\textwidth}
\vskip -2em
\hskip 0.04\textwidth
\begin{minipage}{0.36\textwidth}
  \includegraphics[width=\textwidth]{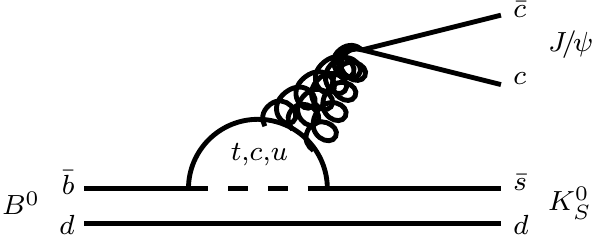}
  \caption{Penguin diagram contributing to \decay{\Bd}{\jpsi\KS}.}\label{Fig:BdPenguin}
\end{minipage}
\end{wrapfigure}
%%%%%%%%%%%%%%%%%%%%%%%%%%%%%%%%%%%%%%%%%%%%%%%%%%%%%%%%%%%%

The SM predictions $\varphi_s=-2\beta_s$ and $\varphi_d=2\beta$ assume 
tree-dominated decays, as in Fig.~\ref{Fig:BdJpsiKs}.
With the precision on some CKM phases reaching the degree level, 
the effects of suppressed penguin topologies (Fig.~\ref{Fig:BdPenguin}) cannot be neglected any 
more~\cite{Fleischer:1999nz,*Fleischer:1999zi,*Fleischer:1999sj,*Faller:2008zc,*DeBruyn:2010hh,Ciuchini:2005mg}. 

Cabibbo-suppressed decay modes, where these topologies are relatively more prominent
can be used to constrain such effects. Ways of using selected measurements
enabling the sizes of penguin amplitudes to be constrained have been described in Ref.~\cite{DeBruyn:2014oga,*DeBruyn:2048174,*Ligeti:2015yma,*Jung:2012mp,*Frings:2015eva}. The LHCb collaboration is pursuing this programme, with studies of the decays 
\decay{\Bs}{\jpsi\KS}~\cite{LHCb-PAPER-2013-015},
\decay{\Bz}{\jpsi\pip\pim}~\cite{LHCb-PAPER-2014-058} and more recently with 
\decay{\Bs}{\jpsi\Kstarzb}~\cite{LHCb-PAPER-2015-034}. 

The measurement of 
$\sin2\beta^\text{eff}$ in the latter mode allows for constraining the 
shift to $\varphi_s^{\cquark\cquarkbar\squark}$ due to penguin topologies
to the range $[-0.018,0.021]$ rad at 68\% CL. Considering the present 
uncertainty of $\pm0.039$ rad, such a shift needs to be to constrained further.

An interesting test of the SM is provided by the measurement of the mixing phase
$\varphi_s^{\squark\squarkbar\squark}$
with a purely penguin-induced mode as \decay{\Bs}{\phi\phi}.
In this case the measured value is $-0.17\pm 0.15\pm 0.03$~\cite{LHCb-PAPER-2014-026},
which is compatible with the SM expectation.

Similarly, the decays \decay{\B}{hh} with $h=\pion,\kaon$ are also sensitive to
penguin topologies (as well as trees) and are sensitive to the CKM phases 
$\gamma$ and $\beta_s$.
LHCb for the first time measured time-dependent \CP-violating observables in
\Bs decays using the decay 
\decay{\Bs}{\Kp\Km}~\cite{LHCb-PAPER-2013-040}. Using methods outlined in 
Refs.~\cite{Fleischer:1999pa,*Fleischer:2007hj,Ciuchini:2005mg}, a combination of this
and other results from \decay{\B}{hh} modes allows to determine 
$-2\beta_s = -0.12 \aerr{0.14}{0.16}$\:\rad
using as input the angle $\gamma$ from tree decays (see below), or 
$\gamma= (63.5\aerr{7.2}{6.7})^\circ$ constraining $-2\beta_s$ to the 
SM value~\cite{LHCb-PAPER-2013-045}.
These values are in principle sensitive to
the amount of U-spin breaking that is allowed in this decay and are given here 
for a maximum allowed breaking of 50\%.

%%%%%%%%%%%%%%%%%%%%%%%%%%%%%%%%%%%%%%%%%%%%%%%%%%%%%%%%%%%%%%%%%%%%%%%%%%%%%%%%%%%%%
\subsection{\boldmath The angle $\gamma$}\label{Sec:Gamma}
The value of $\gamma$ obtained from penguin modes
can be compared to that obtained from tree-dominated
\decay{\B}{\D\kaon} decays, where the \CP-violating phase appears in the interference of
the \decay{\bquark}{\cquark} and \decay{\bquark}{\uquark} topologies. It is
the least precisely known angle of the unitarity triangle, and its determination
from tree decays is considered free from contributions beyond the SM
and unaffected by hadronic uncertainties. Yet its precise determination is important
to test the consistency of the CKM paradigm, and to allow comparisons with determinations
from modes dominated by penguin topologies.

%%%%%%%%%%%%%%%%%%%%%%%
\begin{figure}[b ]
\includegraphics[height=0.4\textwidth]{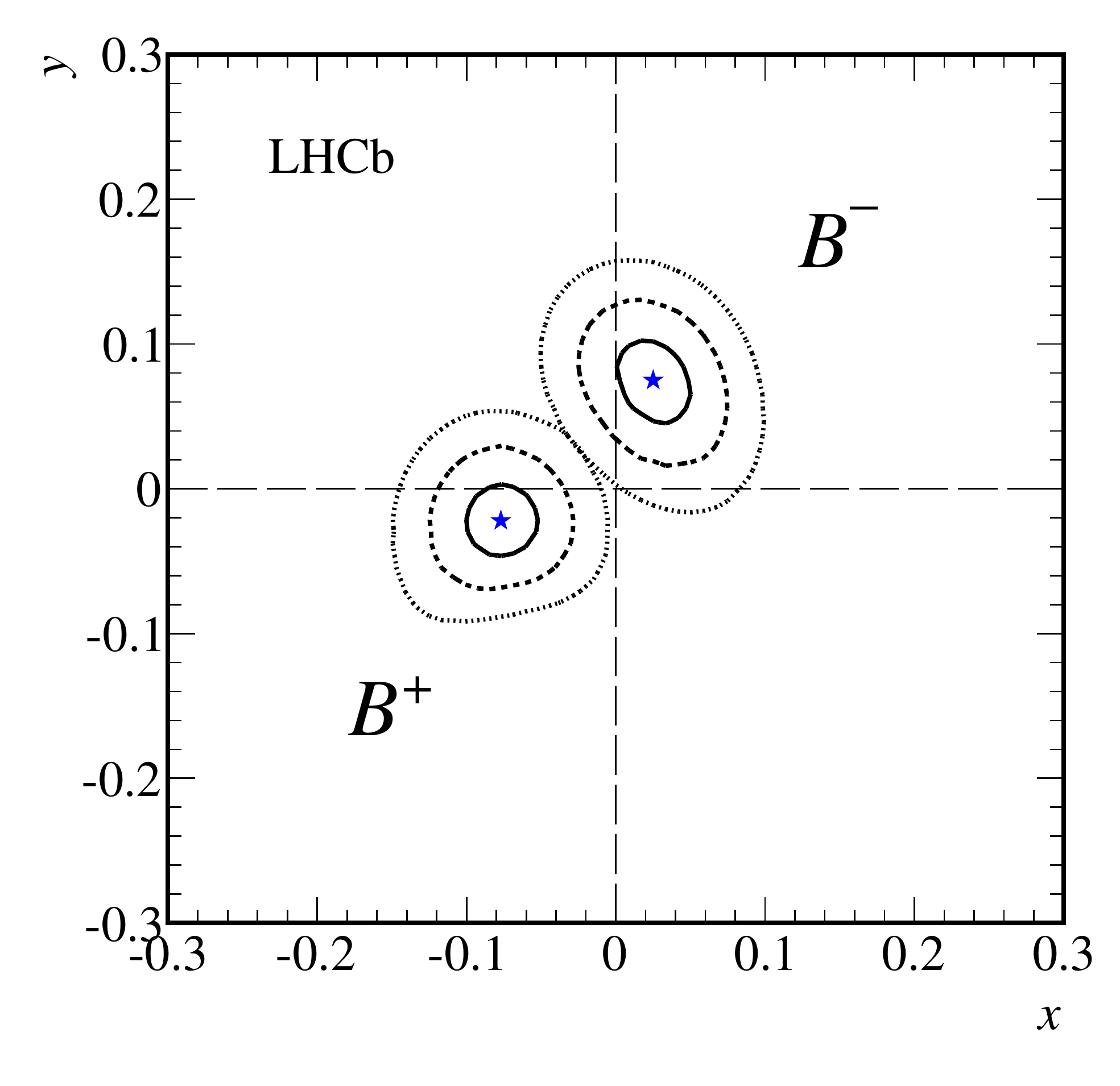}\hskip 0.04\textwidth
\includegraphics[height=0.4\textwidth]{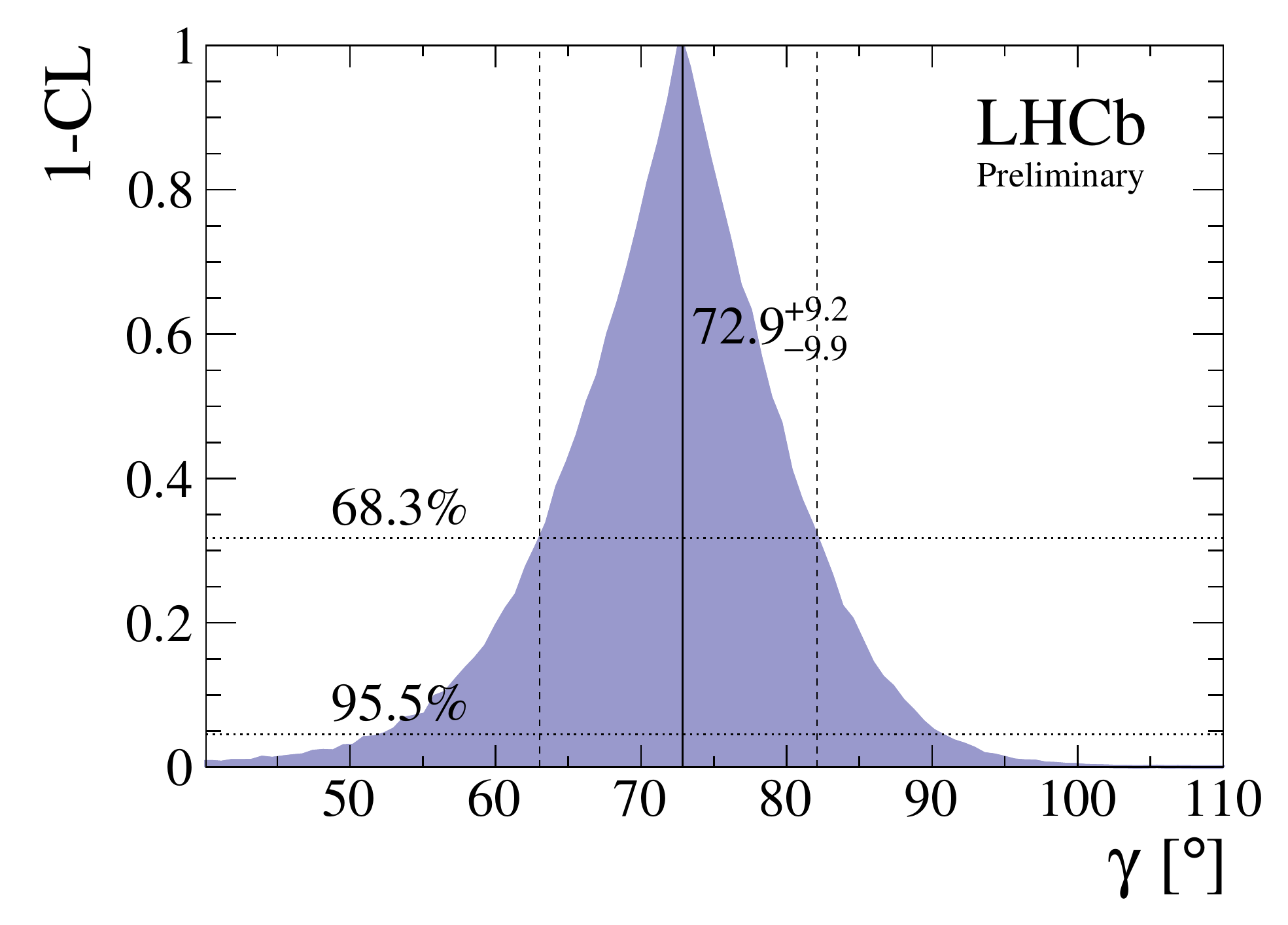}
\caption{(left) Fitted values of $x$ and $y$ for \Bm and 
  \Bp in \decay{\Bpm}{\D\Kpm} with \decay{\D}{\KS h^+h^-} decays~\cite{LHCb-PAPER-2014-041}.
  (right) LHCb combination of \decay{\B}{D\kaon} decays measuring the 
  CKM phase $\gamma$~\cite{LHCb-CONF-2014-004}.}\label{Fig:CONF-2014-004}\label{Fig:2014-041}
\end{figure}
%%%%%%%%%%%%%%%%%%%%%%%
The most precise determination of $\gamma$ from a single decay mode is achieved with 
\decay{\Bp}{D\Kp} followed by \decay{D}{\KS h^+h^-} with 
$h=\pion,\kaon$~\cite{LHCb-PAPER-2014-041}. 
Here the interference of the \Dz and \Dzb decay to $\KS h^+h^-$ 
is exploited to measure \CP asymmetries~\cite{Giri:2003ty}. The method needs external
input in the form of a measurement of the strong phase over the Dalitz plane
of the \D decay, coming from CLEO-c data~\cite{Libby:2010nu}. The 
determined \CP-violating parameters are shown in Fig.~\ref{Fig:2014-041} (left),
and the value of $\gamma$ is $(62\aerr{15}{14})\degrees$. The same decay mode
is also used in a model-dependent measurement~\cite{LHCb-PAPER-2014-017} using
an amplitude model.

An experimentally very different way of determining $\gamma$ is provided by the decay 
\decay{\Bs}{\Dspm\kaon^\mp} \cite{Dunietz:1987bv,*Aleksan:1991nh,*Fleischer:2003yb,LHCb-PAPER-2014-038}. In this case the phase 
is measured in a time-dependent tagged \CP-violation analysis. Using a dataset 
corresponding to 1\:\invfb, LHCb determines $\gamma=(115\aerr{28}{43})\degrees$,
which is not competitive with other methods but will provide important constraints
with more data.

The LHCb $\gamma$ measurements of 
Refs.~\cite{LHCb-PAPER-2014-041,LHCb-PAPER-2014-038,LHCb-PAPER-2012-001,*LHCb-PAPER-2012-055,*LHCb-PAPER-2013-068,*LHCb-PAPER-2014-028} have been
combined into an average value presented at the CKM Workshop~\cite{LHCb-CONF-2014-004}.
Using only \decay{\B}{D\kaon} decay modes one finds $\gamma=(73\aerr{9}{10})\degrees$,
which is more precise than the corresponding combination of measurements from the \B factories~\cite{Bevan:2014iga}. The likelihood profile is shown in Fig.~\ref{Fig:CONF-2014-004} (right). Since then,
new measurements sensitive to $\gamma$ have been performed, notably using the decays 
\decay{\Bm}{\D h^+h^-h^-} with \decay{\D}{h^+h^-}~\cite{LHCb-PAPER-2015-020} and
\decay{\Bm}{\D h^-} with \decay{\D}{h^+h^-\piz}~\cite{LHCb-PAPER-2015-014}. These and other
publications under preparation will be combined in an updated $\gamma$ determination.

\skipit{
%%%%%%%%%%%%%%%%%%%%%%%
\begin{figure}[tb]
\includegraphics[width=0.49\textwidth]{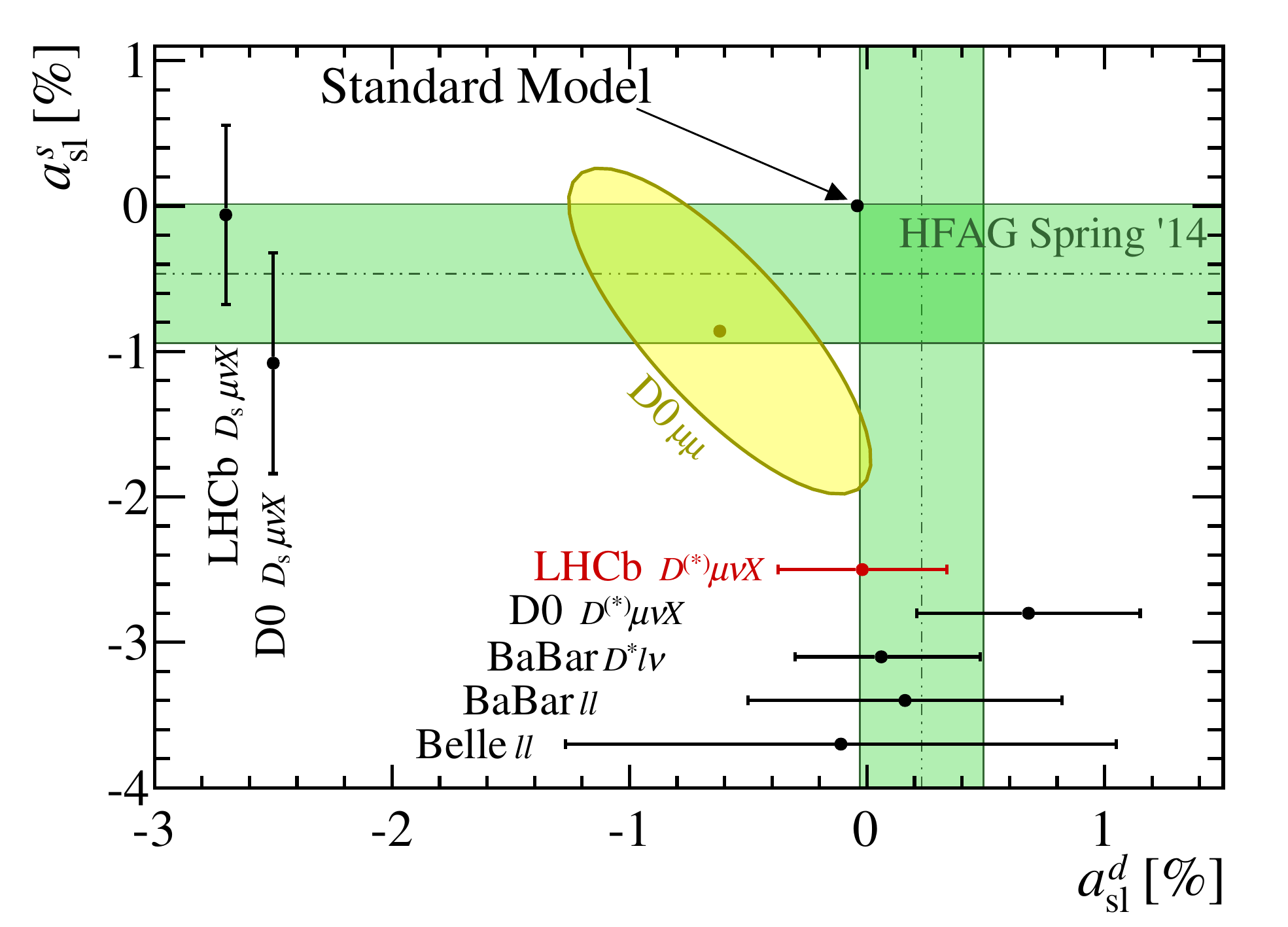}
\caption{ (right) Experimental constraints on $A_\text{sl}^d$ and $A_\text{sl}^s$.}\label{Fig:2014-053}
\end{figure}
%%%%%%%%%%%%%%%%%%%%%%%
The same-sign dimuon asymmetry measured by the \dzero collaboration~\cite{Abazov:2013uma} 
and interpreted as a combination of the semileptonic asymmetries $A_\text{sl}^d$ and $A_\text{sl}^s$ in \Bd and \Bs decays, respectively, remains puzzling. The measured values
differ from the SM expectation by $3\sigma$. So far LHCb has not been able to confirm
or disprove this. The measurement from LHCb follows a different approach, 
looking at the \CP asymmetry
between partially reconstructed \decay{\B}{\Ds\mu\nu} decays, where the flavour of the 
\D identifies that of the \B. The measured value of 
$A_\text{sl}^s$~\cite{LHCb-PAPER-2013-033} and the
newly reported $A_\text{sl}^d$~\cite{LHCb-PAPER-2014-053} are both consistent 
with the SM and the \dzero value. The world average including measurements from the
\B factories and \dzero is not more conclusive. See Fig.~\ref{Fig:2014-053} (right).}

%%%%%%%%%%%%%%%%%%%%%%%%%%%%%%%%%%%%%%%%%%%%%%%%%%%%%%%%%%%%%%%%%%%%%%%%%%%%%%%%%%%%%
\subsection{\boldmath The angle $\alpha$}\label{Sec:alpha}
A precise determination of the third angle of the unitarity triangle, $\alpha=\arg\left(-\frac{\Vtd\Vtbs}{\Vud\Vubs}\right)$ is challenging both at the theoretical and experimental levels. It requires
the study of highly-suppressed \decay{\bquark}{\uquark} transitions, which are affected by
\decay{\bquark}{\dquark} or \decay{\bquark}{\squark} penguin topologies, depending on the chosen
final state. Most promising are angular analyses of \decay{B}{VV} decays, where $V$ is a light
vector meson, typically $\rho$ or \Kstar. At this conference the Belle collaboration
showed a preliminary angular analysis of \decay{\Bz}{\rhop\rhom}~\cite{Vanhoefer}, and 
the LHCb collaboration one of the decay \decay{\Bz}{\rhoz\rhoz}~\cite{LHCb-PAPER-2015-006}.
This latter result solves the tension between the results for the longitudinal polarisation fraction
previously obtained by Babar and Belle~\cite{Aubert:2008au,*Adachi:2012cz}.

%%%%%%%%%%%%%%%%%%%%%%%%%%%%%%%%%%%%%%%%%%%%%%%%%%%%%%%%%%%%%%%%%%%%%%%%%%%%%%%%%%%%%
\subsection{\boldmath The CKM matrix elements \Vub and \Vcb}\label{Sec:Vub}
\begin{figure}[tb]
\includegraphics[height=0.38\textwidth]{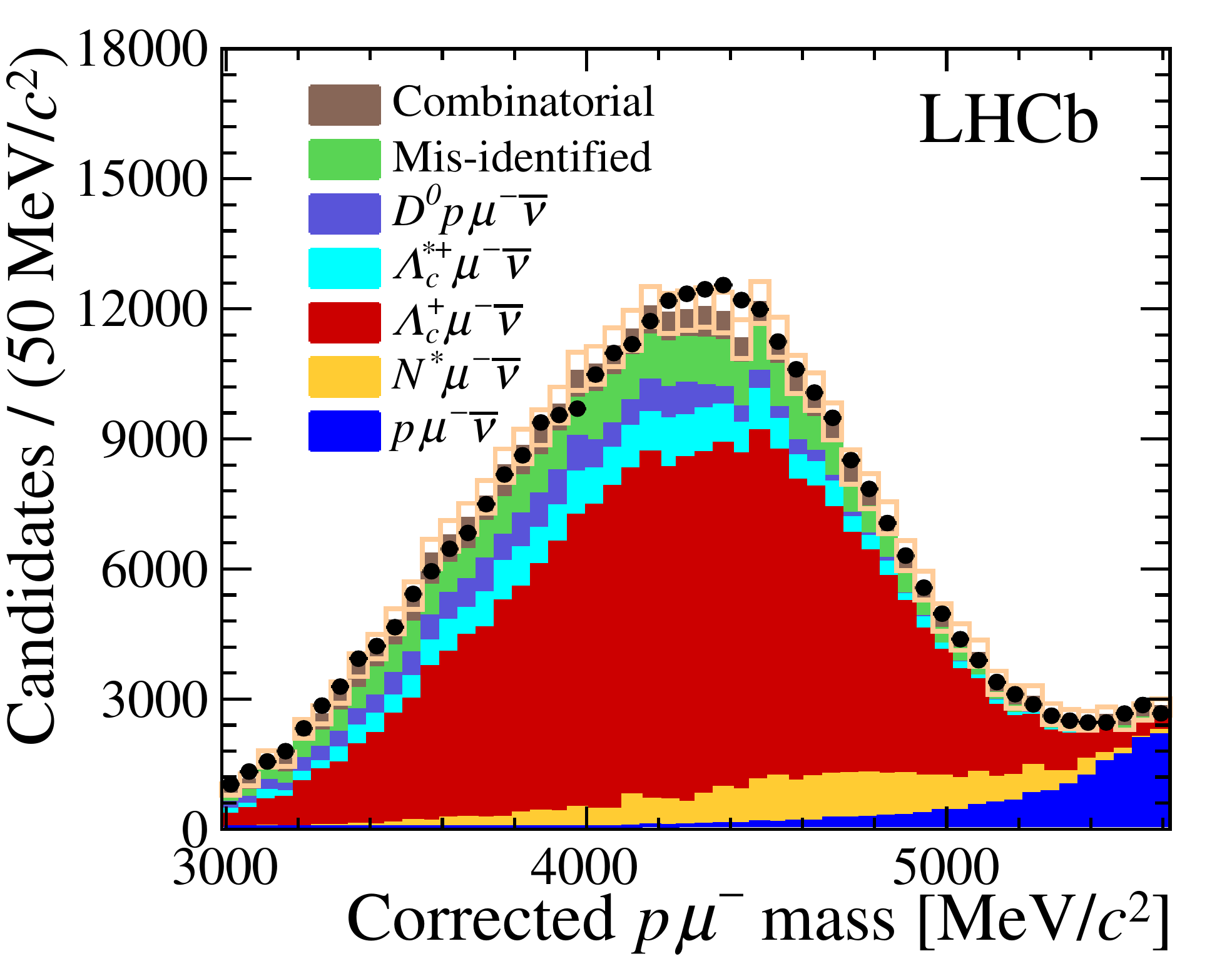}
\includegraphics[height=0.38\textwidth]{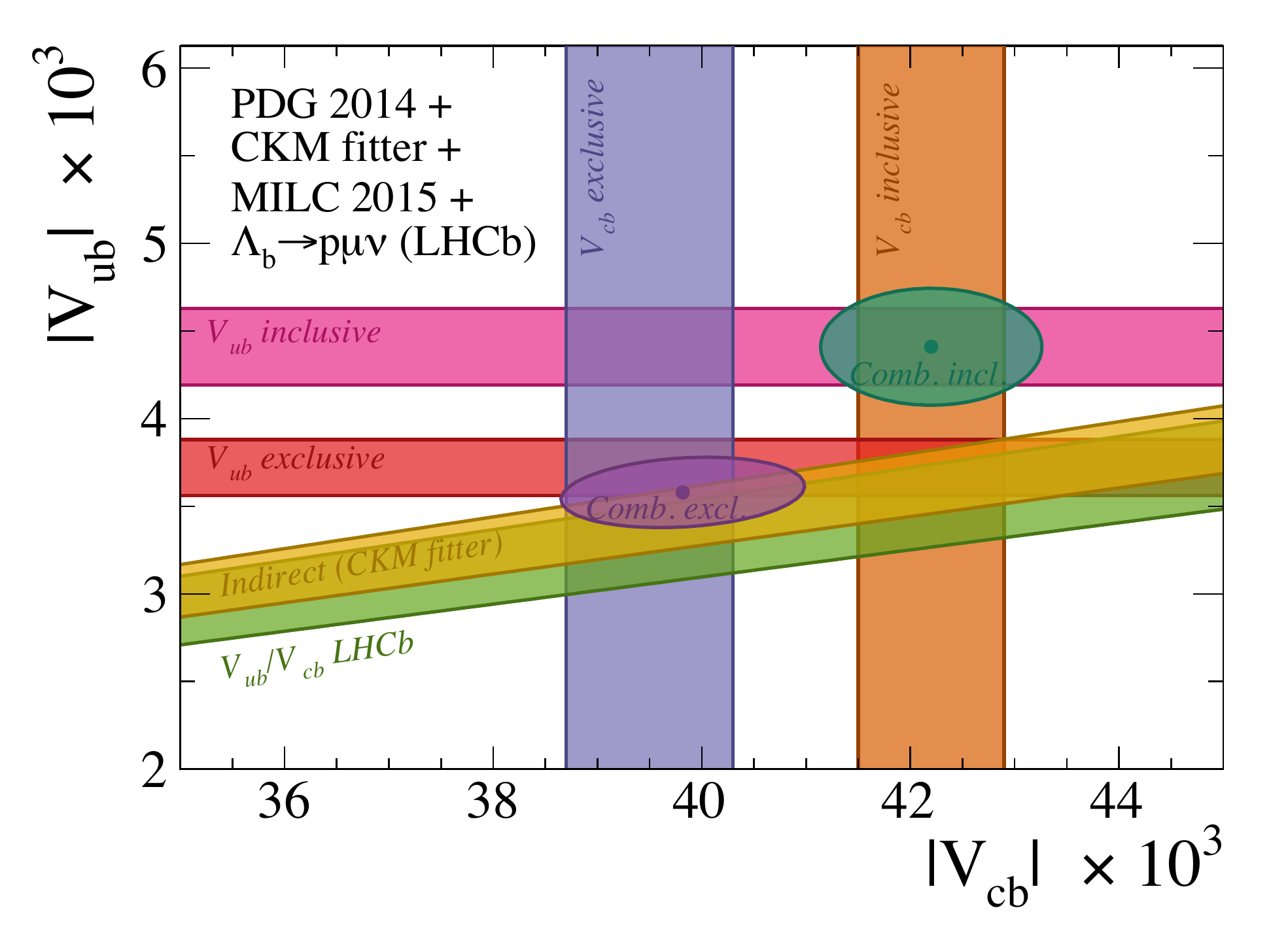}
\caption{(left) Corrected mass spectrum. (right) Experimental measurements of \Vub and \Vcb. Figures from Ref.~\cite{LHCb-PAPER-2015-013}.}\label{Fig:2015-013}
\end{figure}

The determination of the CKM matrix elements \Vub and \Vcb is extremely important for 
closure tests of the UT. It is best measured in semileptonic \decay{\bquark}{(\uquark,\cquark)\ell\nu} decays,
where there are no hadronic uncertainties related to the decay of the emitted \W boson.
Unfortunately, there is a well known tension between the determinations obtained from 
exclusive (\decay{\B}{\pi\ell\nu} for instance) decays and the inclusive mode. It is thus 
important to add new decay modes to the global picture. The LHCb collaboration has done
so by using \Lb baryon decays for the first time \cite{LHCb-PAPER-2015-013}. 
The decay rates of \decay{\Lb}{\proton\mun\nu}
and \decay{\Lb}{\Lc\mun\nu} are compared to determine the ratio $|\Vub|^2/|\Vcb|^2$. It was long thought
impossible to do such an analysis at a hadron collider, due to the lack of a precise neutrino
reconstruction method. At \B factories, fully reconstructing the other \B in the event allows to
unambiguously determine the four-momentum of the missing neutrino. At LHCb, it is the precisely
determined flight path of the \bquark-hadron and the requirement that the \bquark-hadron momentum
must align with its flight direction that allow for the recovery of the neutrino momentum. The
corrected mass $m_\text{corr} = \sqrt{m^2+p_\perp^2}+p_\perp$, the minimal \bquark-hadron 
mass compatible with its direction of flight, is used as discriminating variable. See Fig.~\ref{Fig:2015-013} (left). %%%%%%%%%%%%%%%%%%%%%%%
%%%%%%%%%%%%%%%%%%%%%%%
The two decay rates are turned into a $|\Vub|^2/|\Vcb|^2$ using input from
Lattice calculations~\cite{Detmold:2015aaa}. 

At the same time came an updated
Lattice calculation from the FNAL/MILC collaboration~\cite{Lattice:2015tia} which improves
the determination of $|\Vub|$ from \decay{\Bz}{\pim\ell^+\nu} by the Belle collaboration~\cite{Ha:2010rf}.
Even with these new determinations, the puzzle still remains, as can be seen in Fig.~\ref{Fig:2015-013} (right).

While LHCb only uses muonic decays, the \B factories usually average results from decays 
to muons and electrons under the assumption of lepton universality and do not always give results
separately by lepton flavour (unlike for instance in Ref.~\cite{Aubert:2008yv,*Dungel:2010uk}). This has been discussed in Ref.~\cite{Greljo:2015mma,*Freytsis:2015qca},
in the light of signs of lepton universality violation in 
\decay{\Bp}{\Kp\ellell}~\cite{LHCb-PAPER-2014-024} and \decay{\B}{D^{(*)}(\tau,\ell)\nu} 
decays~\cite{LHCb-PAPER-2015-025,*Lees:2012xj,*Huschle:2015rga}.
Incidentally, the BaBar collaboration presented an inclusive electron spectrum
and a determination of $|\Vub|$ at this conference~\cite{Skovpen}.

%%%%%%%%%%%%%%%%%%%%%%%%%%%%%%%%%%%%%%%%%%%%%%%%%%%%%%%%%%%%%%%%%%%%%%%%%%%%%%%%%%%%%
\subsection{Charmed and Strange Mesons}\label{Sec:Charm}
Searches for \CP violation in charm is a very active field of research,
especially owing to the gigantic charm yields (See Section~\ref{Sec:RunIIj})
collected by the LHCb experiment. Unfortunately, no signs have been found yet and
there is still no clear picture emerging from the measurements of 
the difference between the \CP asymmetries in \decay{\Dz}{\Kp\Km} and 
\decay{\Dz}{\pip\pim}, the world average being $-0.25\pm0.10$~\cite{HFAG}.

New promising results have been shown by the LHCb collaboration. A search for
\CP violation in \decay{\Dz}{\KS\KS} was consistent with zero,
but the improved trigger in Run II will greatly increase the sensitivity of this decay 
mode~\cite{LHCb-PAPER-2015-030}. Secondly, the energy test method was first applied to
particle physics to search for 
local asymmetries in the Dalitz plane of \decay{\Dz}{\pip\pim\piz} 
decays~\cite{LHCb-PAPER-2014-054}.

%%%%%%%%%%%%%%%%%%%%%%%%%%%%%%%%%%%%%%%%%%%%%%%%%%%%%%%%%
\section{First LHCb results from Run II}\label{Sec:RunII}\label{Sec:RunIIc}\label{Sec:RunIIj}
The LHCb collaboration used the LHC 50\:ns ramp-up period of July 2015 to measure the double-differential \jpsi, \jpsi-from-\bquark-hadron~\cite{LHCb-PAPER-2015-037} and charm cross-sections~\cite{LHCb-PAPER-2015-041} at $\sqrt{s}=13\:\tev$. Both measurements were performed directly on triggered candidates using a reduced data format that does not require offline processing.

In the Run~II of the \lhc, a new scheme for the \lhcb software trigger has been introduced, in which the event selection is split into two stages.
This allows for the alignment and calibration to be performed after the first stage of the software trigger
and used directly in the second stage~\cite{Dujany}. The same alignment and calibration are 
propagated offline to ensure consistent reconstruction in the trigger and offline analysis.
The larger timing budget available in the trigger with respect to Run~I also results in the convergence of the online and offline 
track reconstruction, such that offline performance is achieved in the trigger. The near identical performance of the online
and offline reconstruction offers the opportunity to perform physics analyses directly using candidates reconstructed in
the trigger~\cite{LHCb-DP-2012-004}. The analysis described in this paper uses the online reconstruction for the first time in \lhcb. 
The storage of only events containing the triggered candidates leads to an event size that is reduced by an order of magnitude, 
thus permitting an increased event rate with higher efficiency.

Essential to the measurements was a preliminary luminosity calibration using 
beam gas imaging~\cite{FerroLuzzi:2005em}. In absence of a van der Meer scan, neon gas is injected in the LHC vacuum and beam-gas interaction vertices are reconstructed in the vertex detector to reconstruct a three-dimensional image of the two beams and their overlap. A preliminary calibration with an uncertainty of 3.8\% could be determined this way. With a van der Meer scan, more data and more systematic studies, it is expected that this uncertainty will decrease to the level of 1\%, as for the 8\:\tev data~\cite{LHCb-PAPER-2014-047}.

%%%%%%%%%%%%%%%%%%%%%%%
\begin{figure}[tb]
\includegraphics[height=0.35\textwidth]{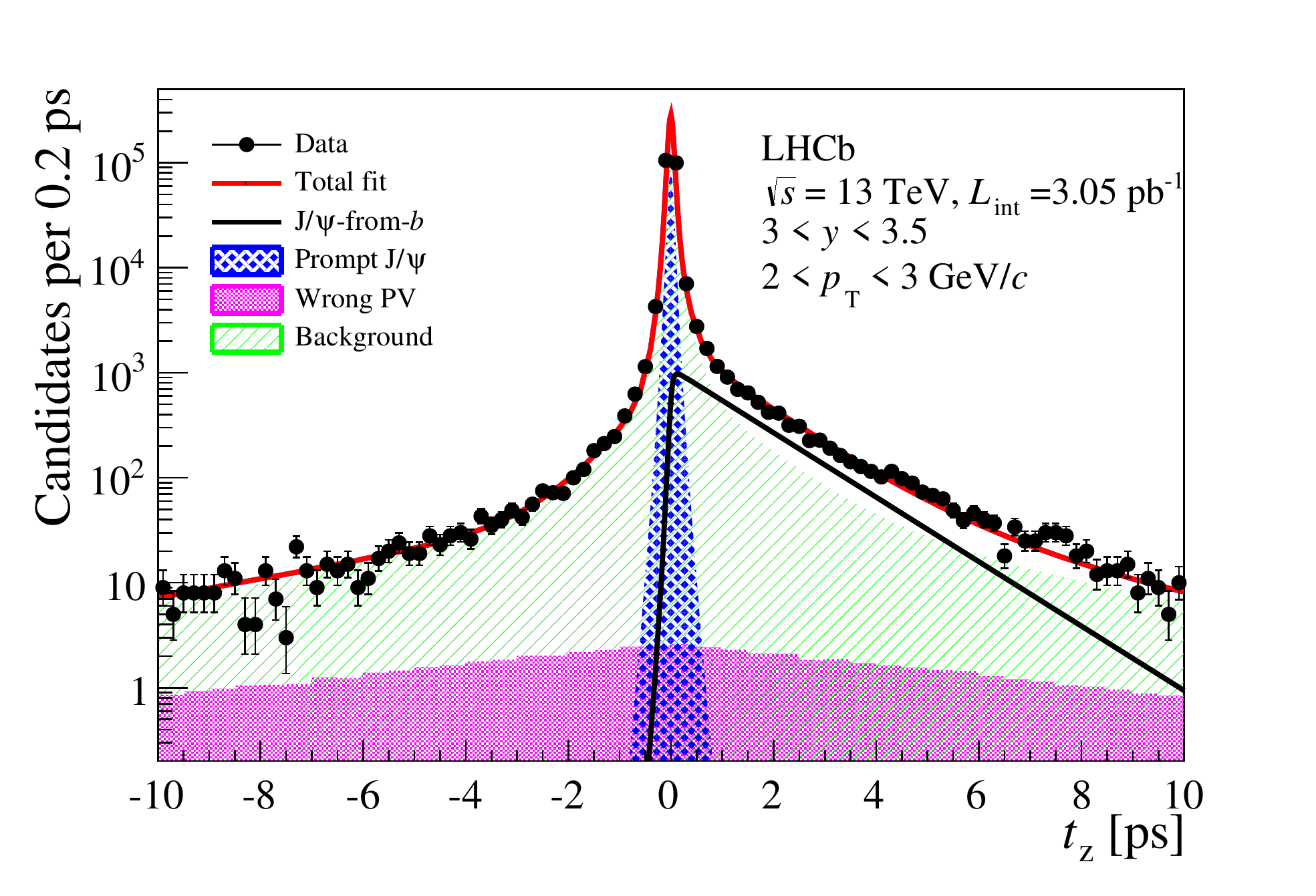}\hskip 0.02
\textwidth\includegraphics[height=0.35\textwidth]{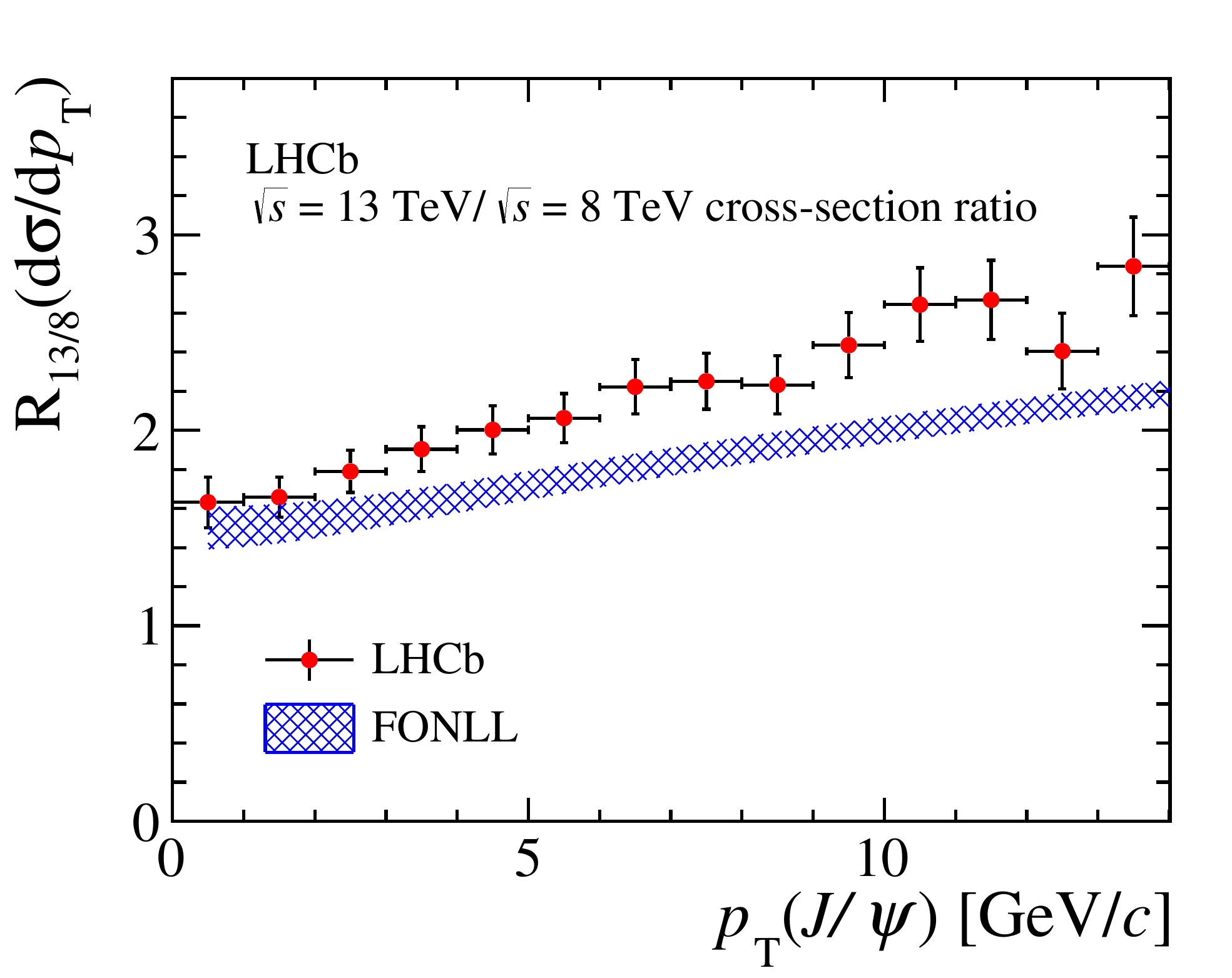}
\caption{(left) \jpsi pseudo-effective lifetime and (right) \pt-dependent cross-section for \jpsi from \bquark-hadrons, compared to expectations from Ref.~\cite{Cacciari:2015fta}. Figures from Ref.~\cite{LHCb-PAPER-2015-037}.}
\label{Fig:2015-037}
\end{figure}
%%%%%%%%%%%%%%%%%%%%%%%
Using $pp$ collision data corresponding to 3\:\invpb of integrated luminosity,
LHCb measured the double differential \jpsi cross-section in bins of \pt and $y$~\cite{LHCb-PAPER-2015-037}. The \decay{\jpsi}{\mumu} signal is further split into \jpsi produced at the primary interaction vertex (prompt) and those produced in decays of \bquark-hadrons. The \jpsi pseudo-effective lifetime $t_z=\frac{(z_\jpsi-z_\text{PV})M_\jpsi}{p_z}$ is used as a proxy for the \bquark-hadron lifetime (Fig.~\ref{Fig:2015-037}, left). While the cross-sections are found in agreement with theoretical expectations within uncertainties, the ratio of $\sqrt{s}=13$ to 8\:\tev ratios, where most uncertainties cancel both in experiment and theory, are found to be slightly above the expectations (Fig.~\ref{Fig:2015-037}, right). The cross-sections are found to be
\begin{align*}\sigma_\text{prompt \jpsi}(\pt<14\:\gevc,2<y<4.5) &= 15.30\pm 0.03\pm 0.86\mub,\\%
 \sigma_\text{\jpsi{}-from-\bquark}(\pt<14\:\gevc,2<y<4.5) &= \phantom{0}2.34\pm 0.01\pm 0.13 \mub.
\end{align*}
Naively extrapolating the latter value from the LHCb fiducial region to the whole 
acceptance using PYTHIA~\cite{LHCb-PROC-2010-056,*Sjostrand:2006za}, 
one gets a \bquark{}\bquarkbar cross-section consistent with 500\:\mub. Incidentally this 
is the value which was used when designing the experiment.

%%%%%%%%%%%%%%%%%%%%%%%
\begin{figure}[tb]
\includegraphics[width=0.48\textwidth]{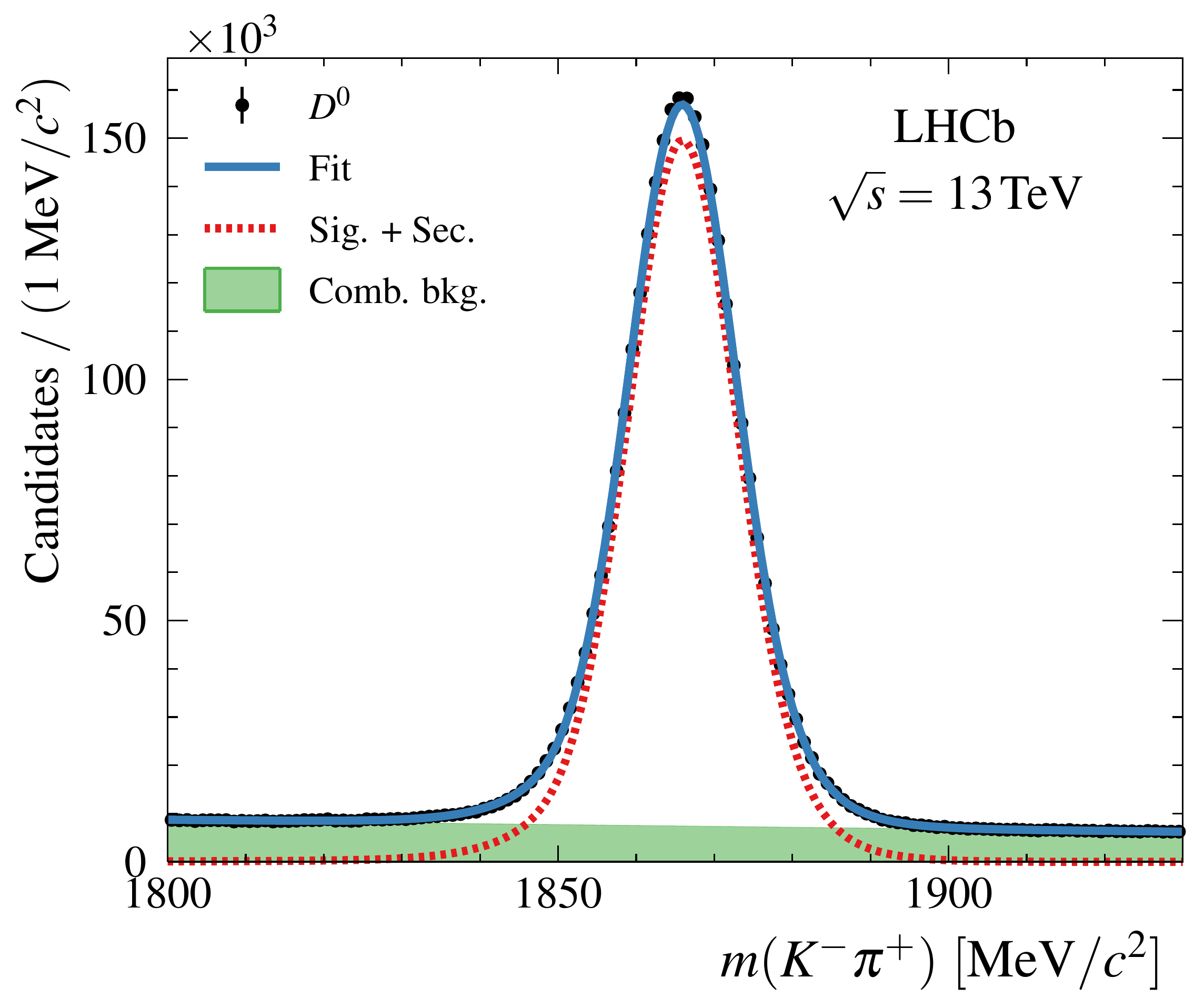}\hskip 0.04\textwidth
\includegraphics[width=0.48\textwidth]{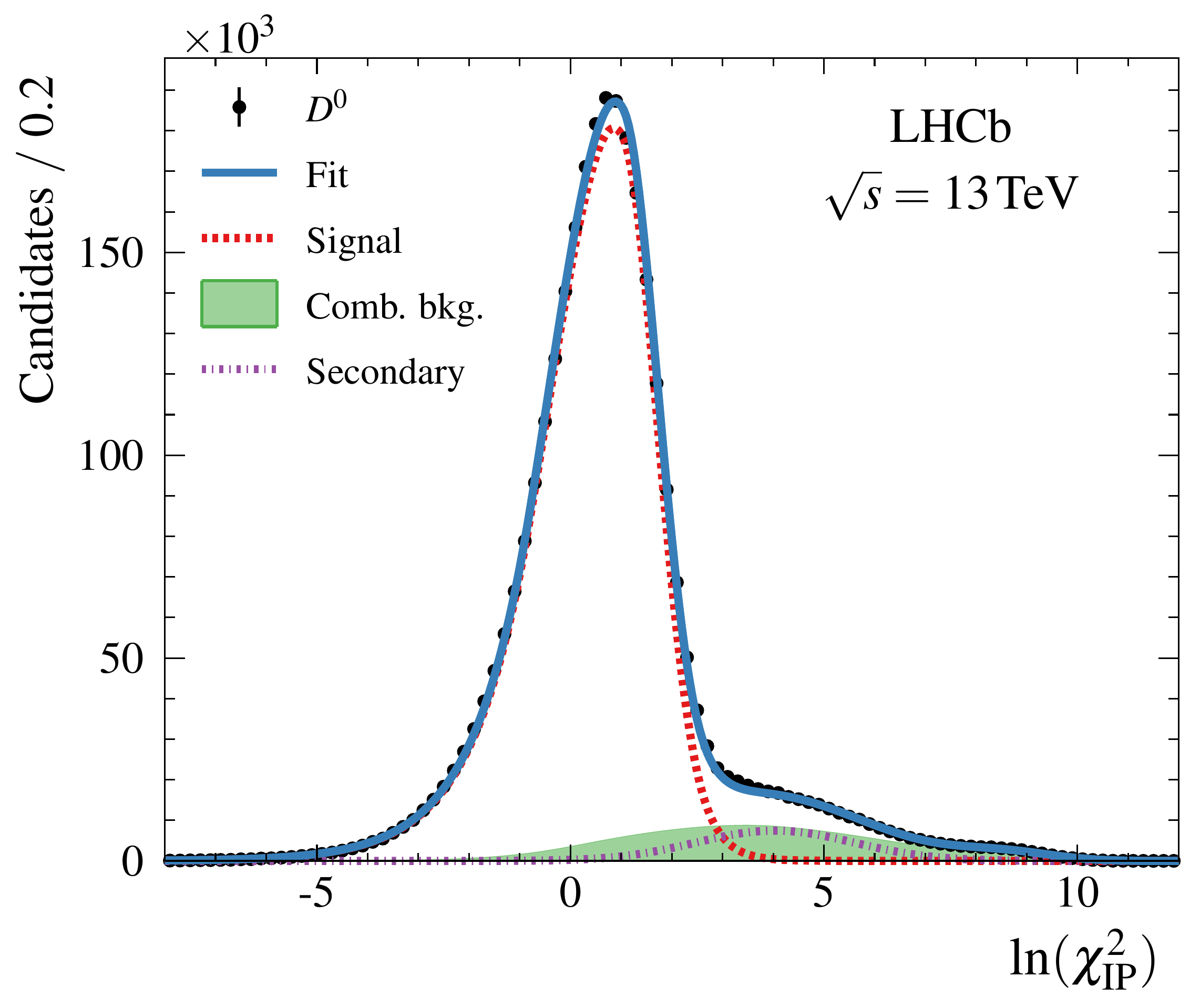}
\caption{\decay{\Dz}{\Km\pip} (left) mass and (right) $\log\chi^2_\text{IP}$ distributions. Figures from Ref.~\cite{LHCb-PAPER-2015-041}.}\label{Fig:2015-041}
\end{figure}
%%%%%%%%%%%%%%%%%%%%%%%
%%%%%%%%%%%%%%%%%%%%%%%
\begin{figure}[tb]
\includegraphics[height=0.22\textheight]{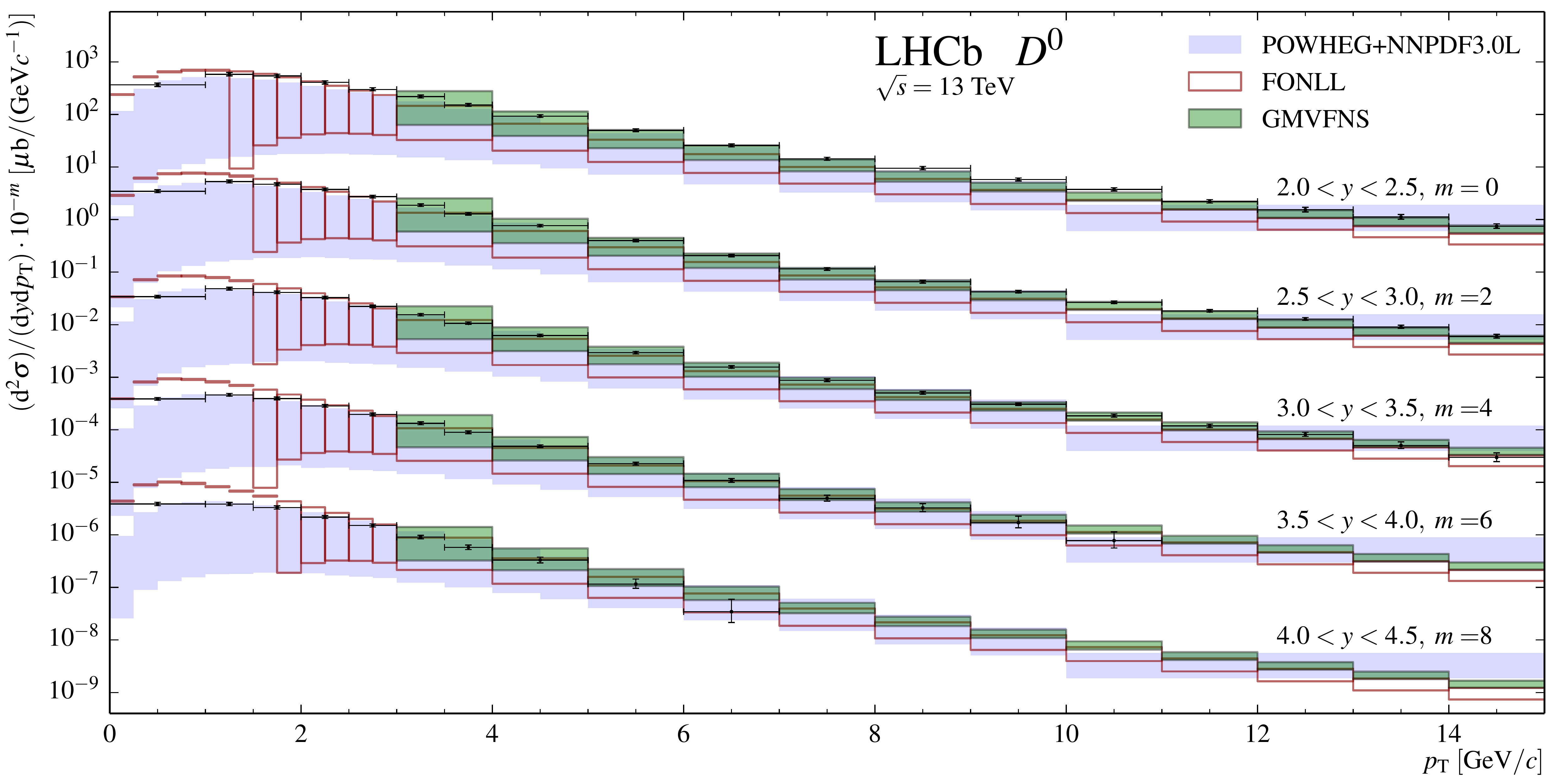}
\includegraphics[height=0.22\textheight]{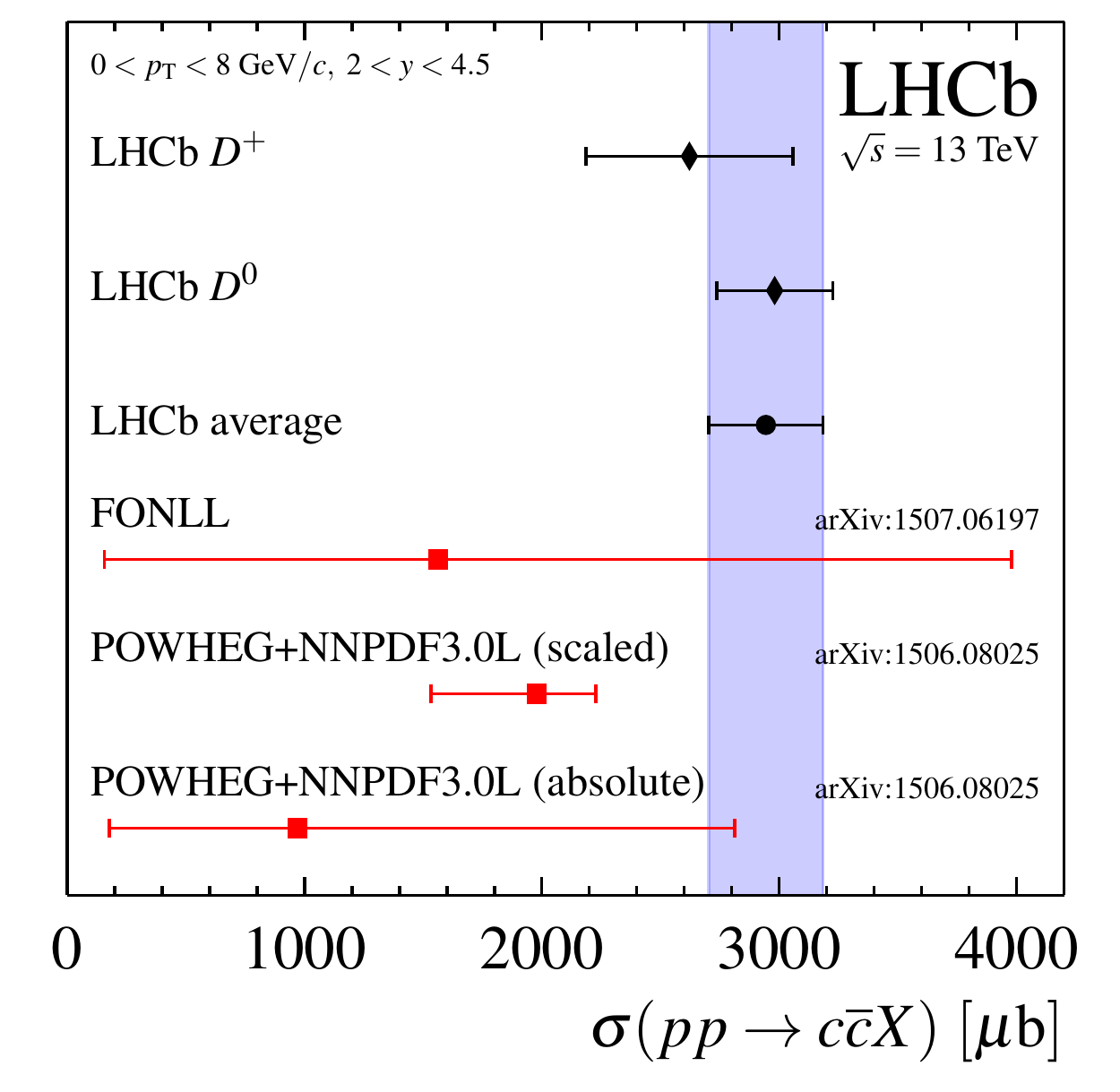}
\caption{(left) \decay{\Dz}{\Km\pip} cross-sections. Each set of measurements and predictions
in a given rapidity bin is offset by a multiplicative factor $10^{-m}$, where the factor $m$ is shown
on the plot. (right) \cquark{}\cquarkbar cross-section in LHCb acceptance from \Dz and \Dp measurements, and comparison with prediction from Refs.~\cite{Gauld:2015yia,Cacciari:2015fta}. The scaled prediction uses the 7\:TeV LHCb measurement~\cite{LHCb-PAPER-2012-041} as input. }
\label{Fig:2015-041b}
\end{figure}
%%%%%%%%%%%%%%%%%%%%%%%
The same dataset and procedure has been used to determine the charm cross-section~\cite{LHCb-PAPER-2015-041}. The double-differential cross-sections for \Dz, \Dp, \Dstar and \Ds mesons are measured using data corresponding to 5\:\invpb. In this case a random trigger was used in the hardware trigger, followed by the full selection in the software trigger. This time the component of charm from \bquark-hadrons is treated as a background and subtracted using the \chisqip (the quantity by which the PV $\chi^2$ increases when the \D meson is included in the fit) as discriminating variable. The mass and $\log\chi^2_\text{IP}$ distributions are shown for \decay{\Dz}{\Km\pip} decays in Fig.~\ref{Fig:2015-041}.

The cross-sections are found to be consistent with predictions from  Refs.~\cite{Gauld:2015yia,Cacciari:2015fta}, but consistently on the high side. An example is shown in Fig.~\ref{Fig:2015-041b}. Using known hadronisation fractions determined at the $\Upsilon$ resonance~\cite{PDG2008} the \cquark{}\cquarkbar cross-section in the LHCb fiducial volume is determined to be 
\begin{align*}
\sigma_\text{prompt \cquark\cquarkbar}(\pt<8\:\gevc,2<y<4.5) &= 
2944 \pm 3 \pm 183 \pm 156 \mub,
\end{align*}
where the uncertainties are statistical, systematic and due to fragmentation fractions.
This value is also compatible with but on the high side of theoretical expectations, which is very encouraging for the LHCb charm physics programme. This number is also relevant for atmospheric neutrino measurements. The charm production cross-section from cosmic rays in the higher atmosphere is a dominant source of uncertainties for measurements of the IceCube collaboration~\cite{Aartsen:2014gkd}. The present measurement at $\sqrt{s}=13\tev$ corresponds to a fixed-target collision with primary cosmic ray energy of 90 PeV.

%%%%%%%%%%%%%%%%%%%%%%%%%%%%%%%%%%%%%%%%%%%%%%%%%%%%%%%%%
\section{Conclusions}
\theabstract
%%%%%%%%%%%%%%%%%%%%%%%%%%%%%%%%%%%%%%%%%%%%%%%%%%%%%%%%%
\section*{Acknowledgements}
The author thanks Vincenzo Vagnoni for the careful reading of the manuscript. Some text is identical to a common previous publication~\cite{2015syce.book...31K}.
%%%%%%%%%%%%%%%%%%%%%%%%%%%%%%%%%%%%%%%%%%%%%%%%%%%%%%%%%
\setboolean{inbibliography}{true} %True once you enter the bibliography
\bibliographystyle{LHCb}    
\bibliography{LHCb-PAPER,LHCb-CONF,LHCb-DP,theory,exp,main,local}

\end{document}